\documentclass[11pt]{article}
\pdfoutput=1

\usepackage{amssymb}
\usepackage{amsmath}
\usepackage{bbold}
\usepackage{color}
\usepackage{colordvi}
\usepackage{dsfont}
\usepackage{fancybox}
\usepackage[footnotesize]{caption2}
\usepackage{graphicx}
\usepackage[center,footnotesize,hang]{subfigure}
\usepackage{bbm}
\usepackage{url}
\usepackage{multirow}
\usepackage{array}
\usepackage{arydshln}
\usepackage{pifont}
\usepackage{mathrsfs}
\usepackage{fancybox}
\usepackage{cite}
\usepackage[colorlinks]{hyperref}
\usepackage{enumitem}
\usepackage{longtable}
\newcommand{\ignore}[1]{}
\usepackage{enumitem}

\newcommand{\PreserveBackslash}[1]{\let\temp=\\#1\let\\=\temp}
\newcolumntype{C}[1]{>{\PreserveBackslash\centering}p{#1}}
\newcolumntype{R}[1]{>{\PreserveBackslash\raggedleft}p{#1}}
\newcolumntype{L}[1]{>{\PreserveBackslash\raggedright}p{#1}}
\allowdisplaybreaks
\newcommand{\bq}{\begin{eqnarray}}
\newcommand{\nq}{\end{eqnarray}}

\usepackage[normalem]{ulem}

\def\bvec#1{\raise1.5ex\hbox{$\rightarrow$}\mkern-16.5mu #1}

\begin{document}
\title{
\begin{flushright}
\hfill\mbox{{\small\tt USTC-ICTS/PCFT-21-29 }} \\[5mm]
\begin{minipage}{0.2\linewidth}
\normalsize
\end{minipage}
\end{flushright}
{\Large \bf
$SO(10)$ models with $A_4$ modular symmetry
\\[2mm]}}
\date{}

\author{
Gui-Jun Ding$^{1,2}$\footnote{E-mail: {\tt
dinggj@ustc.edu.cn}},  \
Stephen~F.~King$^{3}$\footnote{E-mail: {\tt king@soton.ac.uk}}, \
Jun-Nan Lu$^{1, 2}$\footnote{E-mail: {\tt
hitman@mail.ustc.edu.cn}},  \
\\*[20pt]
\centerline{
\begin{minipage}{\linewidth}
\begin{center}
$^1${\it \small
Interdisciplinary Center for Theoretical Study and  Department of Modern Physics,\\
University of Science and Technology of China, Hefei, Anhui 230026, China}\\[2mm]
$^2${\it \small Peng Huanwu Center for Fundamental Theory, Hefei, Anhui 230026, China} \\[2mm]
$^3${\it \small
Physics and Astronomy, University of Southampton, Southampton, SO17 1BJ, U.K.}
\end{center}
\end{minipage}}
\\[10mm]}
\maketitle
\thispagestyle{empty}

\begin{abstract}
We combine $SO(10)$ Grand Unified Theories (GUTs) with $A_4$ modular symmetry and present a comprehensive analysis of the
resulting quark and lepton mass matrices for all the simplest cases.
We focus on the case where the three fermion families in the 16 dimensional spinor representation form a triplet of
$\Gamma_3\simeq A_4$, with a Higgs sector comprising
a single Higgs multiplet $H$ in the ${\mathbf{10}}$ fundamental representation and one Higgs field $\overline{\Delta}$ in the ${\mathbf{\overline{126}}}$ for the minimal models, plus
and one Higgs field $\Sigma$ in the ${\mathbf{120}}$ for the non-minimal models, all with specified modular weights.
The neutrino masses are generated by the type-I and/or type II seesaw mechanisms and
results are presented for each model following an intensive numerical analysis where we have optimized the free parameters of the models in order to match the experimental data.
For the phenomenologically successful models, we present the best fit results in numerical tabular form as well as showing the most interesting graphical correlations between parameters, including leptonic CP phases and neutrinoless double beta decay, which have yet to be measured, leading to definite predictions for each of the models.

\end{abstract}
\newpage

\section{Introduction}

Understanding the pattern of masses and mixings of quarks and leptons (including neutrinos) is one of the greatest challenges in particle physics. The quark and charged lepton masses are described by different interaction strengths with the Higgs doublets within the SM, but their values cannot be predicted. Much effort has been devoted to
addressing the flavour puzzle, and symmetry has been an important guiding principle, where in particular non-Abelian discrete flavor symmetry is particularly suitable to predict the large lepton mixing angles~\cite{King:2013eh}.
However, in conventional flavour symmetry models,
a complicated vacuum alignment is typically required, in which symmetry breaking Higgs fields (flavons) are introduced with their vacuum expectation values (VEVs) oriented along certain directions in flavour space.

Recently modular invariance, inspired by superstring theory with compactified extra dimensions, has been suggested to provide an origin of the flavour symmetry, especially as applied to the neutrino sector~\cite{Feruglio:2017spp}. The finite discrete non-Abelian flavour symmetry groups emerge from the quotient group of the modular group $SL(2,Z)$ over the principal congruence subgroups. The quark and lepton fields transform nontrivially under the finite modular groups and are assigned various modular weights. The modular invariance of Yukawa couplings, whose modular weights do not sum to zero, are necessarily modular forms which are holomorphic functions of the complex modulus $\tau$. In such an approach,
flavon fields other than the modulus may not be needed and the flavour symmetry can be entirely broken by the VEV of the complex modulus field $\tau$. In principle, all higher dimensional operators in the superpotential are completely fixed by modular invariance.

Modular forms are characterised by a positive integer level $N$ and an integer weight $k$ which can be arranged into modular multiplets of the homogeneous finite modular group $\Gamma'_N\equiv\Gamma/\Gamma(N)$~\cite{Liu:2019khw}. They are also organised into modular multiplets of the inhomogeneous finite modular group $\Gamma_N\equiv\overline{\Gamma}/\overline{\Gamma}(N)$ if $k$ is an even number~\cite{Feruglio:2017spp}. The inhomogeneous finite modular group $\Gamma_N$ of the smaller levels
$N=2$~\cite{Kobayashi:2018vbk,Kobayashi:2018wkl,Kobayashi:2019rzp,Okada:2019xqk}, $N=3$~\cite{Feruglio:2017spp,Criado:2018thu,Kobayashi:2018vbk,Kobayashi:2018scp,deAnda:2018ecu,Okada:2018yrn,Kobayashi:2018wkl,Novichkov:2018yse,Nomura:2019jxj,Okada:2019uoy,Nomura:2019yft,Ding:2019zxk,Okada:2019mjf,Nomura:2019lnr,Kobayashi:2019xvz,Asaka:2019vev,Gui-JunDing:2019wap,Zhang:2019ngf,Nomura:2019xsb,Wang:2019xbo,Kobayashi:2019gtp,King:2020qaj,Ding:2020yen,Okada:2020rjb,Nomura:2020opk,Asaka:2020tmo,Okada:2020brs,Yao:2020qyy,Feruglio:2021dte}, $N=4$~\cite{Penedo:2018nmg,Novichkov:2018ovf,deMedeirosVarzielas:2019cyj,Kobayashi:2019mna,King:2019vhv,Criado:2019tzk,Wang:2019ovr,Gui-JunDing:2019wap,Wang:2020dbp,Qu:2021jdy}, $N=5$~\cite{Novichkov:2018nkm,Ding:2019xna,Criado:2019tzk} and $N=7$~\cite{Ding:2020msi} have been analysed and realistic models have been constructed. All the modular forms with integer weights can be generated from the tensor products of weight one modular forms and
the odd weight modular forms which are in representations of $\Gamma'_N$ with $\rho_{\mathbf{r}}(S^2)=-1$. The modular symmetry can not only help to explain the observed lepton mixing patter, but also possibly account for the hierarchical charged lepton masses which can arise from a small deviation from the modular symmetry fixed points~\cite{Okada:2020ukr,Novichkov:2021evw,Feruglio:2021dte}.

These ideas have been generalised in several different directions.
The homogeneous finite modular groups $\Gamma'_N$ provide an even richer structure of modular forms applicable to
flavor model building, with the smaller groups $\Gamma'_3\cong T'$~\cite{Liu:2019khw,Lu:2019vgm}, $\Gamma'_4\cong S'_4$~\cite{Novichkov:2020eep,Liu:2020akv}, $\Gamma'_5\cong A'_5$~\cite{Wang:2020lxk,Yao:2020zml} and $\Gamma'_6\cong S_3\times T'$~\cite{Li:2021buv} having been studied. If the modular weight $k$ of the operator is not an integer, $(c\tau+d)^k$ is not the automorphy factor anymore and some multiplier is needed, consequently the modular group should be extended to its metaplectic covering~\cite{Liu:2020msy}. In this regard,
the formalism of modular invariance has been extended to include the modular forms of rational weights $k/2$~\cite{Liu:2020msy} and $k/5$~\cite{Yao:2020zml}. Furthermore,
generalized CP symmetry can be consistently imposed in the context of modular symmetry, where the modulus is determined to transform as
$\tau\rightarrow-\tau^{*}$ under the action of CP~\cite{Novichkov:2019sqv,Baur:2019kwi,Baur:2019iai,Ding:2021iqp}. The CP transformation matrix is completely fixed by the consistency condition up to an overall phase, it has been shown to coincide with canonical CP transformation in the symmetric basis so that all the coupling constants would be real if the Clebsch-Gordan coefficients in the symmetric basis are real~\cite{Novichkov:2019sqv,Ding:2021iqp}. Finally, the underlying string theory typically involves a compact space with more than one modulus parametrizing its shape, and motivated by this, the modular invariance approach has been extended to incorporate several factorizable~\cite{deMedeirosVarzielas:2019cyj} and non-factorizable moduli~\cite{Ding:2020zxw}.

Grand unified theories (GUTs) are amongst the most well motivated theories beyond the SM, realising the elegant aspiration to unify the three gauge interactions of the SM into a simple gauge group~\cite{Georgi:1974sy}. In GUTs, the quark and lepton fields are embedded into one or few gauge multiplets, resulting in quark and lepton mass matrices whose masses and mixing parameters are related. There is good motivation for imposing a family symmetry together with GUTs in order to address the problem of quark and lepton masses and mixing and especially large lepton mixing~\cite{King:2017guk}. Among the many possible choices of family symmetry, $A_4$ is the minimal choice which admits triplet representations~\cite{Ma:2001dn}, while $SU(5)$~\cite{Georgi:1974sy} is the minimal GUT choice, being the smallest simple group which can accommodate the SM gauge symmetry.
However, combining $A_4$ family symmetry with $SU(5)$ GUTs~\cite{Bjorkeroth:2015ora},
for example, also requires vacuum alignment of the flavons in order to break the $A_4$, and so there is a strong motivation for introducing modular symmetry in such frameworks. Indeed modular symmetry has been
combined with $SU(5)$ GUTs in an $(\Gamma_3\simeq A_4)\times SU(5)$ model
in~\cite{deAnda:2018ecu,Chen:2021zty}. Other modular $SU(5)$ GUT models have also been subsequently constructed based on
$(\Gamma_2\simeq S_3)\times SU(5)$~\cite{Kobayashi:2019rzp,Du:2020ylx}, and $(\Gamma_4\simeq S_4)\times SU(5)$~\cite{Zhao:2021jxg,Ding:2021zbg}.

While $SU(5)$ GUTs is the most minimal choice, it is well known that it does not require neutrino masses,
although they can easily be accommodated as singlet representations of the GUT group.
This motivates the study of the $SO(10)$ GUT group where all known chiral fermions of one generation plus one additional right-handed
neutrino fit into a single 16 dimensional spinor representation of $SO(10)$~\cite{Fritzsch:1974nn}, making neutrino mass inevitable.
Just as in $SU(5)$, so in $SO(10)$ GUTs, there is good motivation
for introducing a family symmetry~\cite{King:2017guk,Bjorkeroth:2017ybg}.
However, just as before, the conventional approach of combining a family symmetry such as $A_4$ with $SO(10)$ GUTs \cite{Grimus:2008tm} also involves the introduction of flavons and vacuum alignment, leading to a complication which can be removed by the use of modular symmetry.
It is perhaps surprising, therefore, that even though there is good motivation for studying modular symmetry as applied to $SO(10)$ GUTs, there seems be no literature on this so far. The purpose of this paper is to remedy this deficiency.

In this paper, we shall perform a comprehensive study of the $\Gamma_3\simeq A_4$ modular symmetry in the framework of supersymmetric (SUSY) $SO(10)$ GUTs. The $SO(10)$ gauge symmetry is spontaneously broken down to the SM gauge group $SU(3)_c\times SU(2)\times U(1)_Y$ by the VEV of Higgs fields $\overline{\Delta}$ in the ${\mathbf{\overline{126}}}$ and/or $\Sigma$ in the ${\mathbf{120}}$. In the minimal $SO(10)$ model, one Higgs multiplet $H$ in the ${\mathbf{10}}$ fundamental representation is required to further break the SM gauge symmetry into $SU(3)_c\times U(1)_{\text{EM}}$, and we shall also assume such a Higgs.
The neutrino masses are generated by the type-I and/or type II seesaw mechanisms~\cite{Minkowski:1977sc,Yanagida:1979as,GellMann:1980vs,Mohapatra:1979ia,Schechter:1980gr} arising from the $SU(2)_L$ singlet and/or triplet components of the Higgs
$\overline{\Delta}$ in the ${\mathbf{\overline{126}}}$ representation.
The purpose of this work is to find the simplest phenomenologically viable $SO(10)$ GUT models based on $A_4$ modular symmetry.
Models we consider are simply defined by organising the three families of fermions in the 16 dimensional spinor representation into a triplet of
$\Gamma_3\simeq A_4$ with a specified modular weight, plus a single Higgs multiplet $H$ in the ${\mathbf{10}}$ fundamental representation, whose modular weight can be taken to be zero without loss of generality, supplemented by one Higgs field $\overline{\Delta}$ in the
${\mathbf{\overline{126}}}$ and/or one Higgs field
$\Sigma$ in the ${\mathbf{120}}$, with specified modular weights. Once these modular weights are specified, the Yukawa couplings are then determined up to a number of overall dimensionless complex coefficients, and the value of the single complex modulus field $\tau$, which is the only flavon in the theory. All models with sums of modular weights up to 10 are considered, and results are presented for each model following an intensive numerical analysis where we have optimized the free parameters of the models in order to match the experimental data. We present results only for the simplest phenomenologically viable models. We find that the minimal models containing only the Higgs fields in the ${\mathbf{10}}$ and the ${\mathbf{\overline{126}}}$ are the least viable, requiring both type I and type II seesaw mechanisms to be present simultaneously and also necessitating at least some sums of modular weights of 10 (if the sums of modular weights were restricted to 8 then no viable such models were found). On the other hand, non-minimal models involving in addition to the Higgs fields in the ${\mathbf{10}}$ and the ${\mathbf{\overline{126}}}$, also a Higgs field in the ${\mathbf{120}}$, proved to be more successful, with many such models being found with sums of modular weights of up to 8 or less, and with the type I seesaw as well as the combined type I+II seesaw also being viable. For the phenomenologically successful models, we present the best fit results in numerical tabular form as well as showing the most interesting graphical correlations between parameters, including leptonic CP phases and the effective mass in neutrinoless double beta decay, which have yet to be measured, leading to definite predictions for each of the models.

This paper is organized as follows. In section~\ref{section2} we briefly review modular symmetry and modular forms of level $N=3$. In section~\ref{section3} we discuss fermion masses from modular forms in SO(10) GUTs, based on the simplest models with three families of fermions in the 16 dimensional spinor representation in a triplet of $\Gamma_3\simeq A_4$, 
plus a single Higgs multiplet $H$ in the ${\mathbf{10}}$ fundamental representation and one Higgs field $\overline{\Delta}$ in the ${\mathbf{\overline{126}}}$ for the minimal models, and plus one Higgs field $\Sigma$ in the ${\mathbf{120}}$ for the non-minimal models, with specified modular weights.
In section~\ref{section4} we present our numerical analysis for the phenomenologically successful models, presenting the best fit results in numerical tabular form as well as showing the most interesting graphical correlations between parameters.
Section~\ref{section5} concludes the paper.

\section{\label{section2}Modular flavor symmetry }

In modular-invariant supersymmetric theories, the action is generally of the form
\begin{equation}
\mathcal{S}=\int d^4xd^2\theta d^2\bar{\theta}\, \mathcal{K}(\tau, \bar{\tau}, \Phi_I, \bar{\Phi}_I)+\int d^4x d^2\theta\, \mathcal{W}(\tau, \Phi_I)+\text{h.c.}\,,
\end{equation}
where $\mathcal{K}(\tau, \bar{\tau}, \Phi_I, \bar{\Phi}_I)$ is the K\"ahler potential, and $\mathcal{W}(\tau, \Phi_I)$ is the superpotential. The supermultiplets are divided into several sectors $\Phi_I$. The action is required to be modular invariant. The modular group $\overline{\Gamma}$ acts on the complex modulus $\tau$ as linear fraction transformation,
\begin{equation}
\tau\rightarrow \gamma\tau=\frac{a\tau+b}{c\tau+d},~~~\text{Im}\tau>0\,,
\end{equation}
where $a$, $b$, $c$ and $d$ are integers satisfying $ad-bc=1$. The modular group $\overline{\Gamma}$ has an infinite number of elements and it can be generated by two generators $S$ and $T$:
\begin{equation}
S:\tau\rightarrow-\frac{1}{\tau}\,,~~~~~~ T:\tau\rightarrow\tau+1\,,
\end{equation}
which obey the relations $S^2=(ST)^3=1$. Under the action of $\overline{\Gamma}$, the supermultiplets $\Phi_I$ transform as~\cite{Feruglio:2017spp}
\begin{equation}
\Phi_I\rightarrow (c\tau+d)^{-k_I}\rho_I(\gamma)\Phi_I\,,
\end{equation}
where $-k_I$ is the modular weight of $\Phi_I$, and $\rho_I$ is a unitary representation of the finite modular group $\Gamma_N=\overline{\Gamma}/\overline{\Gamma}(N)$. $\overline{\Gamma}(N)$ is the principal congruence subgroups of the modular group, and the level $N$ is  fixed in a concrete model. The K\"ahler potential is taken to the minimal form, and it gives rise to kinetic terms of $\Phi_I$ and $\tau$~\cite{Feruglio:2017spp}. The superpotential $\mathcal{W}(\tau, \Phi_I)$ is strongly constrained by modular invariance, and it can be expanded in power series of $\Phi_I$ as follows
\begin{equation}
\mathcal{W}(\tau, \Phi_I)=\sum_n Y_{I_1\ldots I_n}(\tau)\Phi_{I_1}\ldots \Phi_{I_n}\,.
\end{equation}
Modular invariance of $\mathcal{W}$ requires that $Y_{I_1\ldots I_n}(\tau)$ should be a modular forms of weight $k_Y$ and level $N$ transforming in the representation $\rho_{Y}$ of $\Gamma_N$, i.e.,
\begin{equation}
Y_{I_1\ldots I_n}(\tau)\rightarrow Y_{I_1\ldots I_n}(\gamma\tau)=(c\tau+d)^{k_Y}\rho_{Y}(\gamma)Y_{I_1\ldots I_n}(\tau)\,.
\end{equation}
The modular weights and the representations should satisfy the conditions
\begin{eqnarray}
\nonumber&&k_Y=k_{I_1}+\ldots+ k_{I_n}\,,\\
&&\rho_{Y}\otimes \rho_{I_1}\otimes\ldots\otimes\rho_{I_n}\supset \mathbf{1}\,,
\end{eqnarray}
where $\mathbf{1}$ denotes the invariant singlet of $\Gamma_N$.

\subsection{Modular group and modular forms of level 3}
\label{modularforms}

In the present work, we intend to impose modular flavor symmetry in SO(10) GUT to explain the flavor structure of quarks and leptons. We are concerned with the finite modular group $\Gamma_3$ which is isomorphic to $A_4$. $\Gamma_3\cong A_4$ is the even permutation group of four objects, and it can be generated by two generators $S$ and $T$ obeying the previous relations plus the additional one
$T^3=1$,
\begin{equation}
S^2=(ST)^3=T^3=1\,.
\end{equation}
The $A_4$ group has four inequivalent irreducible representations including three one-dimensional representations $\mathbf{1}$, $\mathbf{1}'$, $\mathbf{1}''$ and a three-dimensional representation $\mathbf{3}$. We utilize the symmetric basis in which the generators $S$ and $T$ are represented by unitary and symmetric matrices, i.e.,
\begin{eqnarray}
\nonumber && \mathbf{1}~:~ S=1,~~~~~ T=1 \,,  \\
\nonumber &&\mathbf{1}'~:~ S=1, ~~~~~ T=\omega \,,  \\
\nonumber &&\mathbf{1}''~:~S=1, ~~~~~ T=\omega^{2} \,, \\
&&\mathbf{3}~:~ S=\frac{1}{3}\begin{pmatrix}
    -1~& 2  ~& 2  \\
    2  ~& -1  ~& 2 \\
    2 ~& 2 ~& -1
\end{pmatrix}, ~~~
T=\begin{pmatrix}
    1 ~&~ 0 ~&~ 0 \\
    0 ~&~ \omega ~&~ 0 \\
    0 ~&~ 0 ~&~ \omega^{2}
\end{pmatrix} \,.
\end{eqnarray}
The multiplication rules for the tensor products of the $A_4$ representations are
\begin{eqnarray}
\nonumber&&\mathbf{1}'\otimes\mathbf{1}'=\mathbf{1}'',~~~\mathbf{1}''\otimes\mathbf{1}''=\mathbf{1}',~~~\mathbf{1}'\otimes\mathbf{1}''=\mathbf{1}\,,\\
&&\mathbf{3}\otimes\mathbf{3}=\mathbf{1}\oplus \mathbf{1}'\oplus\mathbf{1}''\oplus\mathbf{3}_S\oplus\mathbf{3}_A\,,
\end{eqnarray}
where $\mathbf{3}_S$ and $\mathbf{3}_A$ denote symmetric and antisymmetric contractions respectively. In terms of the components of the two triplets $\alpha=\left(\alpha_1, \alpha_2, \alpha_3\right)^T$ and $\beta=\left(\beta_1, \beta_2, \beta_3\right)^T$, in this working basis we have
\begin{eqnarray}
\nonumber\begin{pmatrix}
\alpha_1\\
\alpha_2\\
\alpha_3
\end{pmatrix}_{\mathbf{3}}\otimes\begin{pmatrix}
\beta_1\\
\beta_2\\
\beta_3
\end{pmatrix}_{\mathbf{3}}&=&\left(\alpha_1\beta_1+\alpha_2\beta_3+\alpha_3\beta_2\right)_{\mathbf{1}}\oplus\left(\alpha_3\beta_3+\alpha_1\beta_2+\alpha_2\beta_1\right)_{\mathbf{1}'}\oplus\left(\alpha_2\beta_2+\alpha_1\beta_3+\alpha_3\beta_1\right)_{\mathbf{1}''}\\
\label{eq:A4-contraction}&&\oplus\begin{pmatrix}
2\alpha_1\beta_1-\alpha_2\beta_3-\alpha_3\beta_2 \\
2\alpha_3\beta_3-\alpha_1\beta_2-\alpha_2\beta_1\\
2\alpha_2\beta_2-\alpha_1\beta_3-\alpha_3\beta_1
\end{pmatrix}_{\mathbf{3}_S} \oplus \begin{pmatrix}
\alpha_2\beta_3-\alpha_3\beta_2\\
\alpha_1\beta_2-\alpha_2\beta_1\\
\alpha_3\beta_1-\alpha_1\beta_3
\end{pmatrix}_{\mathbf{3}_A}\,.
\end{eqnarray}
The modular forms of level $N$ and weight $k$ span a linear space $\mathcal{M}_k(\Gamma(N))$. The space $\mathcal{M}_k(\Gamma(3))$ has dimension $k+1$, and it can be expressed in terms of the Dedekind eta functions as follows~\cite{schultz2015notes,Liu:2019khw}
\begin{equation}
\mathcal{M}_{k}(\Gamma(3))=\bigoplus_{a+b=k,\,a,b\ge0} \mathbb{C} \frac{\eta^{3a}(3\tau)\eta^{3b}(\tau /3 )}{\eta^k(\tau)}\,,
\end{equation}
where the $\eta(\tau)$ function is
\begin{equation}
\eta(\tau)=q^{1/24}\prod_{n=1}^\infty \left(1-q^n \right)=q^{1/24}\sum^{+\infty}_{n=-\infty} (-1)^nq^{n(3n-1)/2},~~~~q= e^{i 2 \pi\tau}\,.
\end{equation}
The three linearly independent weight 2 modular forms of level 3 can be arranged into a triplet of $A_4$~\cite{Liu:2019khw}:
\begin{equation}
Y^{(2)}_{\mathbf{3}}(\tau)=
\begin{pmatrix}
\varepsilon^2(\tau) \\
\sqrt{2}\,\vartheta(\tau)\varepsilon(\tau) \\
-\vartheta^2(\tau)
\end{pmatrix}\equiv\begin{pmatrix}
Y_1(\tau) \\
Y_2(\tau) \\
Y_3(\tau)
\end{pmatrix}\,,
\end{equation}
with
\begin{equation}
\vartheta(\tau)=3\sqrt{2}\,\frac{\eta^3(3\tau)}{\eta(\tau)},~~~~\varepsilon(\tau)=-\frac{3\eta^3(3\tau)+\eta^3(\tau/3)}{\eta(\tau)}\,.
\end{equation}
Obviously we see that the constraint $Y_2^2+2 Y_1 Y_3=0$ is fulfilled. Notice that $\vartheta(\tau)$ and $\varepsilon(\tau)$ span the linear space of weight 1 modular forms of level 3~\cite{Liu:2019khw}. The expressions of the $q-$expansion of $Y_{1,2,3}(\tau)$ are given by
\begin{eqnarray}
\nonumber&&Y_1(\tau)=1 + 12q + 36q^2 + 12q^3 + 84q^4 + 72q^5+36q^6+96q^7+180q^8+12q^9+216q^{10}+\dots \,, \\
\nonumber&&Y_2(\tau)=-6q^{1/3}\left(1 + 7q + 8q^2 + 18q^3 + 14q^4+31q^5+20q^6+36q^7+31q^8+56q^9+32q^{10} +\dots\right)\,, \\
&&Y_3(\tau)=-18q^{2/3}\left(1 + 2q + 5q^2 + 4q^3 + 8q^4 +6q^5+14q^6+8q^7+14q^8+10q^9+21q^{10}+\dots\right)\,.
\end{eqnarray}
It agrees with the $q-$expansion derived in~\cite{Feruglio:2017spp}, where the modular forms $Y_{1,2,3}(\tau)$ are constructed in terms of $\eta(\tau)$ and its derivative. The whole ring of even weight modular forms can be generated by the modular forms $Y_{1,2,3}(\tau)$ of weight 2. At weight 4, the tensor product of $Y^{(2)}_{\mathbf{3}}$ gives rise to three independent modular multiplets,
\begin{eqnarray}
\nonumber Y^{(4)}_{\mathbf{1}}&=&(Y^{(2)}_{\mathbf{3}}Y^{(2)}_{\mathbf{3}})_{\mathbf{1}}=Y_1^2+2 Y_2 Y_3\,, \\
\nonumber Y^{(4)}_{\mathbf{1}'}&=&(Y^{(2)}_{\mathbf{3}}Y^{(2)}_{\mathbf{3}})_{\mathbf{1}'}=Y_3^2+2 Y_1 Y_2\,,\\
Y^{(4)}_{\mathbf{3}}&=&\frac{1}{2}(Y^{(2)}_{\mathbf{3}}Y^{(2)}_{\mathbf{3}})_{\mathbf{3}_S}=
\begin{pmatrix}
Y_1^2-Y_2 Y_3\\
Y_3^2-Y_1 Y_2\\
Y_2^2-Y_1 Y_3
\end{pmatrix}\,.
\end{eqnarray}
There are seven linearly independent modular forms of level 3 and weight 6 and they decompose as $\mathbf{1}\oplus\mathbf{3}\oplus\mathbf{3}$ under $A_4$,
\begin{eqnarray}
Y^{(6)}_{\mathbf{1}}&=&(Y^{(2)}_{\mathbf{3}}Y^{(4)}_{\mathbf{3}})_{\mathbf{1}}=Y_1^3+Y_2^3+Y_3^3-3 Y_1 Y_2 Y_3\,,\nonumber\\
Y^{(6)}_{\mathbf{3}I}&=&Y^{(2)}_{\mathbf{3}}Y^{(4)}_{\mathbf{1}}=(Y_1^2+2Y_2Y_3)\begin{pmatrix}
Y_1\\
Y_2\\
Y_3
\end{pmatrix}\,,\nonumber\\
\label{eq:MF-w6}Y^{(6)}_{\mathbf{3}II}&=&Y^{(2)}_{\mathbf{3}}Y^{(4)}_{\mathbf{1}'}=
(Y_3^2+2 Y_1Y_2)\begin{pmatrix}
Y_3\\
Y_1\\
Y_2
\end{pmatrix}\,.
\end{eqnarray}
The weight 8 modular forms can be arranged into three singlets $\mathbf{1}$, $\mathbf{1}'$, $\mathbf{1}''$ and two triplets $\mathbf{3}$ of $A_4$,
\begin{eqnarray}
\nonumber Y^{(8)}_{\mathbf{1}}&=&(Y^{(2)}_{\mathbf{3}}Y^{(6)}_{\mathbf{3}I})_{\mathbf{1}}=(Y_1^2+2 Y_2 Y_3)^2\,,\\
\nonumber Y^{(8)}_{\mathbf{1'}}&=&(Y^{(2)}_{\mathbf{3}}Y^{(6)}_{\mathbf{3}I})_{\mathbf{1'}}=(Y_1^2+2 Y_2 Y_3)(Y_3^2+2 Y_1 Y_2)\,, \\
\nonumber Y^{(8)}_{\mathbf{1''}}&=&(Y^{(2)}_{\mathbf{3}}Y^{(6)}_{\mathbf{3}II})_{\mathbf{1''}}=(Y_3^2+2 Y_1 Y_2)^2\,,\\
\nonumber Y^{(8)}_{\mathbf{3}I}&=&Y^{(2)}_{\mathbf{3}}Y^{(6)}_{\mathbf{1}}=(Y_1^3+Y_2^3+Y_3^3-3 Y_1 Y_2 Y_3)\begin{pmatrix}
Y_1 \\
Y_2\\
Y_3
\end{pmatrix}\,,\\
\label{eq:MF-w8}Y^{(8)}_{\mathbf{3}II}&=&(Y^{(2)}_{\mathbf{3}}Y^{(6)}_{\mathbf{3}II})_{\mathbf{3}_A}=(Y_3^2+2 Y_1Y_2)\begin{pmatrix}
Y^2_2-Y_1Y_3\\
Y^2_1-Y_2Y_3\\
Y^2_3-Y_1Y_2
\end{pmatrix}\,.
\end{eqnarray}
The weight 10 modular forms of level 3 decompose as $\mathbf{1}\oplus\mathbf{1}'\oplus\mathbf{3}\oplus\mathbf{3}\oplus\mathbf{3}$ under $A_4$, and they are
\begin{eqnarray}
\nonumber Y^{(10)}_{\mathbf{1}}&=&(Y^{(2)}_{\mathbf{3}}Y^{(8)}_{\mathbf{3}I})_{\mathbf{1}}=(Y_1^2+2 Y_2 Y_3)(Y_1^3+Y_2^3+Y_3^3-3 Y_1 Y_2 Y_3)\,,\\
\nonumber Y^{(10)}_{\mathbf{1'}}&=&(Y^{(2)}_{\mathbf{3}}Y^{(8)}_{\mathbf{3}I})_{\mathbf{1'}}=(Y_3^2+2 Y_1 Y_2)(Y_1^3+Y_2^3+Y_3^3-3 Y_1 Y_2 Y_3)\,, \\
\nonumber Y^{(10)}_{\mathbf{3}I}&=&Y^{(2)}_{\mathbf{3}}Y^{(8)}_{\mathbf{1}}=(Y_1^2+2 Y_2 Y_3)^2\begin{pmatrix}
Y_1 \\
Y_2\\
Y_3
\end{pmatrix}\,,\\
\nonumber Y^{(10)}_{\mathbf{3}II}&=&Y^{(2)}_{\mathbf{3}}Y^{(8)}_{\mathbf{1}'}=(Y_1^2+2 Y_2 Y_3)(Y_3^2+2 Y_1 Y_2)\begin{pmatrix}
Y_3 \\
Y_1\\
Y_2
\end{pmatrix}\,,\\
\label{eq:MF-w10}Y^{(10)}_{\mathbf{3}III}&=&Y^{(2)}_{\mathbf{3}}Y^{(8)}_{\mathbf{1}''}=(Y_3^2+2 Y_1 Y_2)^2\begin{pmatrix}
Y_2 \\
Y_3\\
Y_1
\end{pmatrix}\,.
\end{eqnarray}
We summarize the even weight modular forms of level 3 and their decomposition under $A_4$ in table~\ref{Tab:Level3-MF}.

\begin{table}[t!]
\centering
\begin{tabular}{|c|c|}
\hline  \hline

Modular weight $k$ & Modular form $Y^{(k)}_{\mathbf{r}}$ \\ \hline

$k=2$ & $Y^{(2)}_{\mathbf{3}}$\\ \hline

$k=4$ & $Y^{(4)}_{\mathbf{1}}\,,~ Y^{(4)}_{\mathbf{1}'}\,,~ Y^{(4)}_{\mathbf{3}}$\\ \hline

$k=6$ & $Y^{(6)}_{\mathbf{1}}\,,~ Y^{(6)}_{\mathbf{3}I}\,,~ Y^{(6)}_{\mathbf{3}II}$\\ \hline

$k=8$ & $Y^{(8)}_{\mathbf{1}}\,,~ Y^{(8)}_{\mathbf{1}'}\,,~ Y^{(8)}_{\mathbf{1}''}\,,~ Y^{(8)}_{\mathbf{3}I}\,,~ Y^{(8)}_{\mathbf{3}II}$\\ \hline

$k=10$ & $Y^{(10)}_{\mathbf{1}}\,,~ Y^{(10)}_{\mathbf{1}'}\,,~ Y^{(10)}_{\mathbf{3}I}\,,~ Y^{(10)}_{\mathbf{3}II}\,,~ Y^{(10)}_{\mathbf{3}III}$ \\ \hline \hline
\end{tabular}
\caption{\label{Tab:Level3-MF} Modular forms of level 3 up to weight 10, where the superscripts indicate the modular weights and the subscripts denote how they transform under the $A_4$ modular symmetry. Here $Y^{(6)}_{\mathbf{3}I}$ and $Y^{(6)}_{\mathbf{3}II}$ stand for the two linearly independent weight 6 modular forms in the triplet $\mathbf{3}$ of $A_4$. Similar notations are adopted for $Y^{(8)}_{\mathbf{3}I}$, $Y^{(8)}_{\mathbf{3}II}$ and $Y^{(10)}_{\mathbf{3}I}$, $Y^{(10)}_{\mathbf{3}II}$, $Y^{(10)}_{\mathbf{3}III}$. }
\end{table}

\section{Fermion masses from modular forms in SO(10) GUTs }
\label{section3}

The $SO(10)$ GUT theory embeds all SM fermions of a generation plus a right-handed neutrino into a single spinor representation denoted as $\mathbf{16}$. As a consequence, the gauge sector and the fermionic matter sector are generally quite simple. However, the same is not true of the Higgs sector. Since the larger GUT symmetry group $SO(10)$ needs to be broken down to the Standard Model gauge group $SU(3)_C\times SU(2)_L\times U(1)_Y$, generally one needs to introduce a large number of Higgs multiplets, with different symmetry properties under gauge transformations.
There are two options when constructing $SO(10)$ GUT models, one can use Higgs multiplets in low-dimensional representations of $SO(10)$ then the Lagrangian contains non-renormalizable terms, or the Higgs multiplet fields are in high-dimensional representations and the Lagrangian is renormalizable.
In this work, we shall stick to the renormalizable Supersymmetric (SUSY) $SO(10)$ models  for simplicity.
The fermion masses are generated by the Yukawa couplings of fermion bilinears in the spinor representation $\mathbf{16}$ with the Higgs fields multiplets. From the following tensor product of $SO(10)$ group
\begin{equation}
16 \otimes 16 = 10_{S} \oplus 120_{A} \oplus 126_{S}\,,
\end{equation}
where the subscripts $S$ and $A$ stand for the symmetric and antisymmetric parts of the tensor products respectively in flavor space, we see that the Higgs fields in the $SO(10)$ representations $\mathbf{10}$, $\mathbf{120}$ and $\overline{\mathbf{126}}$ can have renormalizable Yukawa couplings. Hence the Yukawa superpotential in renormalizable $SO(10)$ models can be generally written as follows:
\begin{equation}
\label{eq:Yukawa-supp}\mathcal{W}_Y=\mathcal{Y}_{ab}^{10} \psi_{a} \psi_{b}H+\mathcal{Y}_{ab}^{\overline{126}}  \psi_{a} \psi_{b}\overline{\Delta}+ \mathcal{Y}_{ab}^{120}\psi_{a} \psi_{b} \Sigma\,,
\end{equation}
where $a,b=1, 2, 3$ are indices of generation, $\psi$ refers to the matter fields in the $\mathbf{16}$ dimensional representation of the $SO(10)$, $H$, $\overline{\Delta}$ and $\Sigma$ denote the Higgs fields in the representations $\mathbf{10}$, $\overline{\mathbf{126}}$ and $\mathbf{120}$ respectively. The Yukawa coupling matrices $\mathcal{Y}^{10}$ and $\mathcal{Y}^{\overline{126}}$ are $3\times3$ complex symmetric in flavor space, while $\mathcal{Y}^{120}$ is a complex antisymmetric matrix, i.e.
\begin{equation}
\mathcal{Y}_{ab}^{10} = \mathcal{Y}_{ba}^{10},\quad \mathcal{Y}_{ab}^{\overline{126}} = \mathcal{Y}_{ba}^{\overline{126}}, \quad
\mathcal{Y}_{ab}^{120} = -\mathcal{Y}_{ba}^{120}\,.
\end{equation}
After the $SO(10)$ GUT symmetry breaking by the vacuum expectation values of the Higgs fields, all the above Yukawa couplings with $H$, $\overline{\Delta}$ and $\Sigma$ contribute to the masses of both quarks and charged leptons, and in particularly the Yukawa couplings of $\overline{\Delta}$ provide the Majorana mass terms of both right-handed and left-handed neutrinos. Therefore in general the effective mass matrix of light neutrino receives contributions from type-I and type-II seesaw mechanisms.

Under the Pati-Salam group $SU(4) \times SU(2)_{L} \times SU(2)_{R}$, the relevant $SO(10)$ representations have the following decomposition
\begin{eqnarray}
16 & = & (4,2,1) + (\overline{4},1,2)\\
10 & = & (6,1,1) + (1,2,2)\\
120 & = & (15,2,2) + (6,3,1) + (6,1,3) + (1,2,2) + (10,1,1) + (\overline{10},1,1)\\
126 & = & (10,1,3) + (\overline{10},3,1) + (15,2,2) + (6,1,1)
\end{eqnarray}
where the components $(15,2,2)$ and $(1,2,2)$ both contain
a pair of the $Y = \pm 1$ $SU(2)_{L}$ Higgs doublets, whose neutral components give masses to the fermions. The component $(10,1,3)$ contained in
$\overline{126}$ gives Majorana masses of the right-handed neutrinos and the Majorana masses of the left-handed neutrinos are generated due to the $(\overline{10},3,1)$ component of $\overline{126}$.
 To be more specific, the $SO(10)$ Higgs fields $H$, $\overline{\Delta}$ and $\Sigma$ have SM Higgs components, and the decomposition of theses fields under the SM gauge symmetry $SU(3)_C\times SU(2)_L\times U(1)_Y$ are
\begin{eqnarray}
\nonumber&&10\supset(\mathbf{1}, \mathbf{2}, 1/2)\oplus(\mathbf{1}, \mathbf{2}, -1/2)\equiv \Phi^{10}_d\oplus\Phi^{10}_u\,,\\
\nonumber&&120\supset(\mathbf{1}, \mathbf{2}, 1/2)\oplus(\mathbf{1}, \mathbf{2}, -1/2)\oplus(\mathbf{1}, \mathbf{2}, 1/2)\oplus(\mathbf{1}, \mathbf{2}, -1/2)\equiv \Phi^{120}_d\oplus\Phi^{120}_u\oplus\Phi'^{120}_d\oplus\Phi'^{120}_u\,,\\
&&\overline{126}\supset(\mathbf{1}, \mathbf{2}, 1/2)\oplus(\mathbf{1}, \mathbf{2}, -1/2)\oplus(\mathbf{1}, \mathbf{1}, 0)\oplus(\mathbf{1}, \mathbf{3}, 1)\equiv \Phi^{\overline{126}}_d\oplus\Phi^{\overline{126}}_u\oplus \overline{\Delta}_R\oplus \overline{\Delta}_L
\end{eqnarray}
Decomposing the Yukawa superpotential in Eq.~\eqref{eq:Yukawa-supp}, we find the quark and lepton mass matrices are of the following form
\begin{eqnarray}
\nonumber&&M_u=v^{10}_u\mathcal{Y}^{10}+v^{\overline{126}}_u\mathcal{Y}^{\overline{126}}+(v^{120}_u+v'^{120}_u)\mathcal{Y}^{120}\,,\\
\nonumber&&M_d=v^{10}_d\mathcal{Y}^{10}+v^{\overline{126}}_d\mathcal{Y}^{\overline{126}}+(v^{120}_d+v'^{120}_d)\mathcal{Y}^{120}\,,\\
\nonumber&&M_{\ell}=v^{10}_d\mathcal{Y}^{10}-3v^{\overline{126}}_d\mathcal{Y}^{\overline{126}}+(v^{120}_d-3v'^{120}_d)\mathcal{Y}^{120}\,,\\
\nonumber&&M_D=v^{10}_u\mathcal{Y}^{10}-3v^{\overline{126}}_u\mathcal{Y}^{\overline{126}}+(v^{120}_u-3v'^{120}_u)\mathcal{Y}^{120}\,,\\
\nonumber&&M_R=v^{\overline{126}}_R\mathcal{Y}^{\overline{126}}\\ \label{eq:Yukawa}
&&M_L=v^{\overline{126}}_L\mathcal{Y}^{\overline{126}}\,,
\end{eqnarray}
where the VEVs are defined as
\begin{eqnarray}
\nonumber&&v^{10}_{u,d}=\langle\Phi^{10}_{u,d}\rangle\,,~~~v^{120}_{u,d}=\langle\Phi^{120}_{u,d}\rangle\,,~~~v'^{120}_{u,d}=\langle\Phi'^{120}_{u,d}\rangle\,,\\
&&v^{\overline{126}}_{u,d}=\langle\Phi^{\overline{126}}_{u,d}\rangle\,,~~~v^{\overline{126}}_R=\langle\overline{\Delta}_R\rangle\,,~~~v^{\overline{126}}_L=\langle\overline{\Delta}_L\rangle\,.
\end{eqnarray}
Therefore both type-I and type-II seesaw mechanisms contribute to the neutrino masses, and the effective mass matrix of light neutrinos is given by
\begin{equation}
\label{eq:mnu-I-II-seesaw}M_{\nu}=M_{L}-M_DM^{-1}_RM^T_D\,,
\end{equation}
where the first and the second terms denote the type-II and a type-I seesaw contributions respectively, the two contributions to neutrino mass depend on two different parameters $v_L$ and $v_R$. It is possible to have a symmetry breaking pattern in SO(10) such that the first contribution (the type-II term) dominates over the type-I term. It is convenient to redefine the parameters as follows
\begin{eqnarray}
\nonumber&& \mathcal{\widetilde{Y}}^{10}=\frac{v^{10}_u}{v_u}\mathcal{Y}^{10},~~~\mathcal{\widetilde{Y}}^{\overline{126}}=\frac{v^{\overline{126}}_d}{v^{10}_d}\frac{v^{10}_u}{v_u}\mathcal{Y}^{\overline{126}},~~~\mathcal{\widetilde{Y}}^{120}=\frac{v^{120}_d+v'^{120}_d}{v^{10}_d}\frac{v^{10}_u}{v_u}\mathcal{Y}^{120}    \\
\nonumber&& r_1=\frac{v_u}{v_d}\frac{v^{10}_d}{v^{10}_u}
,~~~r_2=\frac{v^{\overline{126}}_u}{v^{\overline{126}}_d}\frac{v^{10}_d}{v^{10}_u},~~~r_3=\frac{v^{120}_u+v'^{120}_u}{v^{120}_d+v'^{120}_d}\frac{v^{10}_d}{v^{10}_u}\,,
\\
\nonumber&&c_e=\frac{v^{120}_d-3v'^{120}_d}{v^{120}_d+v'^{120}_d},~~~c_{\nu}=\frac{v^{120}_u-3v'^{120}_u}{v^{120}_d+v'^{120}_d}\frac{v^{10}_d}{v^{10}_u}\,,\\
\label{eq:Higgs-VEVs}&&v_R=v^{\overline{126}}_R\frac{v^{10}_d}{v^{\overline{126}}_d}\frac{v_u}{v^{10}_u},~~~v_L=v^{\overline{126}}_L\frac{v^{10}_d}{v^{\overline{126}}_d}\frac{v_u}{v^{10}_u}
\end{eqnarray}
where $v_u$ and $v_d$ are the VEVs of the MSSM Higgs pair $H_u$ and $H_d$. The parameters $r_a$ ($a=1,2,3$ and $c_b$ ($b=e,\nu$) are the mixing parameters which relate the $H_{u,d}$ to the doublets in the various GUT
multiplets~\cite{Dutta:2005ni,Grimus:2006bb}. Notice that $\mathcal{\widetilde{Y}}^{10}$, $\mathcal{\widetilde{Y}}^{\overline{126}}$ and $\mathcal{\widetilde{Y}}^{120}$ are proportional to the Yukawa matrices $\mathcal{Y}^{10}$, $\mathcal{Y}^{\overline{126}}$ and $\mathcal{Y}^{120}$ respectively, and the coefficients $\frac{v^{10}_u}{v_u}$, $\frac{v^{\overline{126}}_d}{v^{10}_d}\frac{v^{10}_u}{v_u}$ and $\frac{v^{120}_d+v'^{120}_d}{v^{10}_d}\frac{v^{10}_u}{v_u}$ can be absorbed into the coupling constants $\alpha_i$, $\beta_i$ and $\gamma_i$ given in Eq.~\eqref{eq:W_Y}. Then the mass matrices of the quarks and leptons can be written as
\begin{eqnarray}
M_u &=&\left(\mathcal{\widetilde{Y}}^{10} + r_2 \mathcal{\widetilde{Y}}^{\overline{126}} +r_3 \mathcal{\widetilde{Y}}^{120}\right)v_u,\nonumber\\
M_d &=& r_1 \left(\mathcal{\widetilde{Y}}^{10}+ \mathcal{\widetilde{Y}}^{\overline{126}} + \mathcal{\widetilde{Y}}^{120}\right)v_d\,, \nonumber\\
M_{\ell} &=& r_1\left(\mathcal{\widetilde{Y}}^{10}-3 \mathcal{\widetilde{Y}}^{\overline{126}} + c_e \mathcal{\widetilde{Y}}^{120}\right)v_d\,, \nonumber\\
M_{D} &=&\left(\mathcal{\widetilde{Y}}^{10}-3 r_2 \mathcal{\widetilde{Y}}^{\overline{126}} + c_\nu \mathcal{\widetilde{Y}}^{120}\right)v_u, \nonumber\\
\label{eq:mass-matrices-para}M_R&=&v_R\mathcal{\widetilde{Y}}^{\overline{126}},~~~M_L=v_L\mathcal{\widetilde{Y}}^{\overline{126}} \,.
\label{yuk}
\end{eqnarray}
The effective neutrino mass matrix is still given by Eq.~\eqref{eq:mnu-I-II-seesaw}.

\subsection{Combining SO(10) with $A_4$ modular symmetry}

The three generation of fermions $\psi_{1,2,3}$ are assumed to transform as a triplet $\mathbf{3}$ under $A_4$ modular symmetry, and its modular weight is denoted as $k_F$. All the Higgs multiplets $H$, $\overline{\Delta}$ and $\Sigma$ are assigned to trivial $A_4$ singlet with modular weights $k_{10}$, $k_{\overline{126}}$ and $k_{120}$ respectively. Without loss of generality we can set $k_{10}=0$, since $k_{10}$ can be absorbed
by shifting the modular weight of the matter fields. Modular invariance requires the Yukawa couplings $\mathcal{Y}_{ab}^{10}$,  $\mathcal{Y}_{ab}^{\overline{126}}$ and $\mathcal{Y}_{ab}^{120}$ in Eq.~\eqref{eq:Yukawa-supp} are just modular forms of level 3. The most general Yukawa superpotential invariant under both $SO(10)$ and  modular symmetry is of the following form
\begin{eqnarray}
\nonumber\mathcal{W}_{Y}&=&\sum_{\mathbf{r}_a}
\alpha_a\left((\psi\psi H)_{\mathbf{r}'_a}Y_{\mathbf{r}_a}^{(2k_F+k_{10})}(\tau)\right)_{\mathbf{1}}+\sum_{\mathbf{r}_c}
\beta_b\left((\psi\psi \Sigma)_{\mathbf{r}'_b}Y_{\mathbf{r}_b}^{(2k_F+k_{120})}(\tau)\right)_{\mathbf{1}}\\ \label{eq:W_Y}
&&~~~+\sum_{\mathbf{r}_c}\gamma_c\left((\psi\psi \overline{\Delta})_{\mathbf{r}'_c}Y_{\mathbf{r}_c}^{(2k_F+k_{\overline{126}})}(\tau)\right)_{\mathbf{1}}\,,
\end{eqnarray}
where the representations $\mathbf{r}_{a,b,c}$ and $\mathbf{r}'_{a,b,c}$ fulfill $\mathbf{r}'_a\otimes \mathbf{r}_a=\mathbf{r}'_b\otimes \mathbf{r}_b=\mathbf{r}'_c\otimes \mathbf{r}_c=\mathbf{1}$, and the allowed values of $\mathbf{r}_{a,b,c}$ depend on the modular weights of matter fields and Higgs multiplets, as summarized in table~\ref{Tab:Level3-MF}.
Note that for simplicity we have written $\alpha_a\equiv \alpha_{\mathbf{r}_a}^{(2k_F+k_{10})}$,
$\beta_b\equiv \beta_{\mathbf{r}_b}^{(2k_F+k_{120})}$, $\gamma_c\equiv \gamma_{\mathbf{r}_c}^{(2k_F+k_{\overline{126}})}$.
Using the group theory contraction rules of $A_4$ in Eq.~\eqref{eq:A4-contraction}, the Yukawa matrices can be straightforwardly derived
in terms of the complex coefficients $\alpha_a, \beta_b, \gamma_c$ and the
modular forms $Y_{\mathbf{r}}^{(k)}(\tau)$ introduced in subsection~\ref{modularforms}.
For low values of modular weights, we have
\begin{eqnarray}
\nonumber 2k_F+k_{10}=0~&:&~ \mathcal{Y}^{10}=\alpha_1 S^{(0)}_{\mathbf1}\,,\\
\nonumber 2k_F+k_{10}=2~&:&~  \mathcal{Y}^{10}=\alpha_1 S^{(2)}_{\mathbf3}\,,\\
\nonumber 2k_F+k_{10}=4~&:&~  \mathcal{Y}^{10}= \alpha_1 S^{(4)}_{\mathbf1}+\alpha_2 S^{(4)}_{\mathbf1'}+\alpha_3 S^{(4)}_{\mathbf3}\,,\\
\nonumber 2k_F+k_{10}=6~&:&~  \mathcal{Y}^{10}=\alpha_1 S^{(6)}_{\mathbf1} + \alpha_2 S^{(6)}_{\mathbf{3}I}+\alpha_3 S^{(6)}_{\mathbf{3}II}\,,\\
\nonumber 2k_F+k_{10}=8~&:&~  \mathcal{Y}^{10}= \alpha_1 S^{(8)}_{\mathbf1}+\alpha_2 S^{(8)}_{\mathbf1'}+\alpha_3 S^{(8)}_{\mathbf1''}+\alpha_4 S^{(8)}_{\mathbf{3}I}+\alpha_5 S^{(8)}_{\mathbf{3}II}\,,\\ \label{eq:Y_10_126}
2k_F+k_{10}=10~&:&~ \mathcal{Y}^{10}= \alpha_1 S^{(10)}_{\mathbf1}+\alpha_2 S^{(10)}_{\mathbf1'} + \alpha_3 S^{(10)}_{\mathbf{3}I}+\alpha_4 S^{(10)}_{\mathbf{3}II}+\alpha_5 S^{(10)}_{\mathbf{3}III}\,,
\end{eqnarray}
where $S^{(k)}_{\mathbf r}$ are symmetric $3\times 3$ matrices defined below.
The Yukawa matrix $\mathcal{Y}^{\overline{126}}$ is of a similar form with $\alpha_i$ replaced by $\gamma_i$. By comparison $\mathcal{Y}^{120}$ is antisymmetric, it only receives contribution from the antisymmetric contraction $\left(\psi\psi\right)_{\mathbf{3}_A}$, and its concrete forms crucially depends on the modular weight $k=2k_F+k_{120}$ with the antisymmetric matrices $A^{(k)}_{\mathbf r}$:
\begin{eqnarray}
\nonumber 2k_F+k_{120}=2~&:&~\mathcal{Y}^{120}=\beta_1 A^{(2)}_{\mathbf3}\,,\\
\nonumber 2k_F+k_{120}=4~&:&~\mathcal{Y}^{120}=\beta_1 A^{(4)}_{\mathbf3} \,,\\
\nonumber 2k_F+k_{120}=6~&:&~\mathcal{Y}^{120}=\beta_1 A^{(6)}_{\mathbf{3}I} +\beta_2 A^{(6)}_{\mathbf{3}II}\,,\\
\nonumber 2k_F+k_{120}=8~&:&~\mathcal{Y}^{120}=\beta_1 A^{(8)}_{\mathbf{3}I} +\beta_2 A^{(8)}_{\mathbf{3}II}\,,\\ \label{eq:Yukawa_Y_120}
2k_F+k_{120}=10~&:&~\mathcal{Y}^{120}=\beta_1 A^{(10)}_{\mathbf{3}I} + \beta_2 A^{(10)}_{\mathbf{3}II} + \beta_3 A^{(10)}_{\mathbf{3}III}\,.
\end{eqnarray}
For convenience we have defined the symmetric and antisymmetric $3\times 3$ matrices
\begin{eqnarray}
S^{(k)}_{\mathbf1} =
Y_{\mathbf1}^{(k)}(\tau)\left(
\begin{matrix}
 1 ~& 0 ~& 0 \\
 0 ~& 0 ~& 1 \\
 0 ~& 1 ~& 0 \\
\end{matrix}
\right)\,,~~~
S^{(k)}_{\mathbf1'} =
Y_{\mathbf1'}^{(k)}(\tau)\left(
\begin{matrix}
 0 ~& 0 ~& 1 \\
 0 ~& 1 ~& 0 \\
 1 ~& 0 ~& 0 \\
\end{matrix}
\right)\,,~~~
S^{(k)}_{\mathbf1''} =
Y_{\mathbf1''}^{(k)}(\tau)\left(
\begin{matrix}
 0 ~& 1 ~& 0 \\
 1 ~& 0 ~& 0 \\
 0 ~& 0 ~& 1 \\
\end{matrix}
\right)\,,
\end{eqnarray}
and
\begin{eqnarray}
\label{eq:Sk-Ak}S^{(k)}_{{\mathbf3}} =
\left(\begin{matrix}
 2Y^{(k)}_{{{\mathbf3}},1}(\tau) ~&~ -Y^{(k)}_{{{\mathbf3}},3}(\tau)  ~&~ -Y^{(k)}_{{{\mathbf3}},2}(\tau) \\
 -Y^{(k)}_{{{\mathbf3}},3}(\tau) ~&~ 2Y^{(k)}_{{{\mathbf3}},2}(\tau)  ~&~ -Y^{(k)}_{{{\mathbf3}},1}(\tau) \\
 -Y^{(k)}_{{{\mathbf3}},2}(\tau) ~&~ -Y^{(k)}_{{{\mathbf3}},1}(\tau)  ~&~ 2Y^{(k)}_{{{\mathbf3}},3}(\tau)
\end{matrix}
\right)\,,~~
A^{(k)}_{{\mathbf3}} =
\left(\begin{matrix}
 0 ~&~ Y^{(k)}_{{{\mathbf3}},3}(\tau)  ~&~ -Y^{(k)}_{{{\mathbf3}},2}(\tau) \\
 -Y^{(k)}_{{{\mathbf3}},3}(\tau) ~&~ 0  ~&~ Y^{(k)}_{{{\mathbf3}},1}(\tau) \\
 Y^{(k)}_{{{\mathbf3}},2}(\tau) ~&~ -Y^{(k)}_{{{\mathbf3}},1}(\tau)  ~&~ 0
\end{matrix}\right)\,,
\end{eqnarray}
which depend on the modular forms $Y_{\mathbf{r}}^{(k)}(\tau)$ introduced in subsection~\ref{modularforms}.
Notice that there are no non-trivial modular forms of weight zero, the expressions of $S^{(6)}_{\mathbf{3}I,II}$, $S^{(8)}_{\mathbf{3}I,II}$ and  $S^{(10)}_{\mathbf{3}I,II,III}$ can be read out from Eq.~\eqref{eq:Sk-Ak} with the components of triplet modular forms in Eqs.~(\ref{eq:MF-w6}, \ref{eq:MF-w8}, \ref{eq:MF-w10}), and similarly for $A^{(6)}_{\mathbf{3}I,II}$, $A^{(8)}_{\mathbf{3}I,II}$ and  $A^{(10)}_{\mathbf{3}I,II,III}$.

\begin{table}[t!]
\centering
\begin{tabular}{|c|c|c|c|} \hline  \hline
Parameters & $\mu_i\pm1\sigma$ & Parameters & $\mu_i\pm1\sigma$ \\ \hline
$m_{t}/\text{GeV}$ & $83.155 \pm 3.465$ & $\theta_{12}^{q}$ & $0.229 \pm 0.001$ \\
$m_{b}/\text{GeV}$ & $0.884 \pm 0.035$ & $\theta_{13}^{q}$ & $0.0037 \pm 0.0004$ \\
  $m_{u}/m_{c}$ & $0.0027 \pm 0.0006$ & $\theta_{23}^{q}$ & $0.0397 \pm 0.0011 $ \\
  $m_{c}/m_{t}$ & $0.0025 \pm 0.0002$ & $\delta_{CP}^{q}/^{\circ}$ & $56.34 \pm 7.89 $ \\
  $m_{d}/m_{s}$ & $0.051 \pm 0.007$ & $\sin^{2}\theta_{12}^{l}$ & $0.304\pm 0.012$ \\
  $m_{s}/m_{b}$ & $0.019 \pm 0.002$ & $\sin^{2}\theta_{23}^{l}$ & $0.573_{-0.020}^{+0.016}$ \\
  $m_{e}/m_{\mu}$ & $0.0048 \pm 0.0002$ & $\sin^{2}\theta_{13}^{l}$ & $0.02219_{-0.00063}^{+0.00062}$ \\
  $m_{\mu}/m_{\tau}$ & $0.059 \pm 0.002$ & $\delta_{CP}^{l}/^{\circ}$ & $197_{-24}^{+27}$ \\
$m_{b}/m_{\tau}$ & $0.73 \pm 0.03$ & $r\equiv\Delta m_{21}^{2}/\Delta m_{31}^{2}$ & $0.02948 \pm 0.00087$\\
\hline \hline
\end{tabular}
\caption{\label{Tab:parameter_values} The best fit values $\mu_i$ and $1\sigma$ uncertainties of the quark and lepton parameters when evolve to the GUT scale as calculated in~\cite{Ross:2007az}, with the SUSY breaking scale $M_{\text{SUSY}}=500$ GeV and $\tan\beta=10$, where the error widths represent $1\sigma$ intervals.
The parameter $r\equiv \Delta m_{21}^{2}/ \Delta m_{31}^{2}$ is the ratio of neutrino mass-squared differences. The values of lepton mixing angles, leptonic Dirac CP violation phases $\delta^{l}_{CP}$ and the neutrino mass squared difference are taken from NuFIT 5.0~\cite{Esteban:2020cvm}.
}
\end{table}

\subsection{Minimal models}

If only one of the Higgs fields $H$, $\overline{\Delta}$, $\Sigma$ is employed, all fermion mass matrices are proportional to the same Yukawa matrix which can be diagonalized by a basis transformation. As a consequence, there would be no flavor mixing between up type and down type quarks or between charged leptons and neutrinos. Hence at least two Higgs fields are necessary for realistic fermion spectrum and flavor mixing. One of the Higgs fields must be a $\overline{\Delta}$ in the $SO(10)$ representation $\overline{\mathbf{126}}$ in order to generate tiny neutrino masses. The Yukawa couplings $\mathcal{Y}_{ab}^{\overline{126}}$ of $\overline{\Delta}$ to the fermions provide Majorana masses to both right-handed and left-handed neutrinos and the seesaw mechanism is naturally realized, as can be seen from Eq.~\eqref{eq:mass-matrices-para}. In the so-called minimal $SO(10)$ GUT, only the Higgs fields $H$ and $\overline{\Delta}$ couple to the fermion sector. It is remarkable that the Yukawa superpotential is completely fixed by the summation of the modular weights of the matter and Higgs fields. In the following we present three benchmark models.

\begin{itemize}[labelindent=-1.5em, leftmargin=1.6em]

\item{\texttt{Minimal model 1:} $(2k_F+k_{10}\,, 2k_F+k_{\overline{126}}) = (10,6)$}
\\
Without loss of generality, the modular weight of $H$ can be taken to be vanishing, i.e., $k_{10}=0$. Then we have $k_F=5$ and $k_{\overline{126}}=-4$. Modular invariance requires that the modular forms of weight 6 and weight 10 should be present in the Yukawa interactions, and the most general form of the superpotential reads as
\begin{eqnarray}
\nonumber\mathcal{W}_Y &=&
\alpha_1 Y_{\mathbf{1}}^{(10)} \psi \psi H
+\alpha_2 Y_{\mathbf{1}'}^{(10)} \psi \psi H
+\alpha_3 Y_{\mathbf{3}I}^{(10)} \psi \psi H
+\alpha_4 Y_{\mathbf{3}II}^{(10)} \psi \psi H
+\alpha_5 Y_{\mathbf{3}III}^{(10)} \psi \psi H \\ \label{eq:SP_minimal_model_1}
&&+\gamma_1 Y_{\mathbf{1}}^{(6)} \psi \psi \overline{\Delta}
+\gamma_2 Y_{\mathbf{3}I}^{(6)} \psi \psi \overline{\Delta}
 +\gamma_3 Y_{\mathbf{3}II}^{(6)} \psi \psi \overline{\Delta}\,.
\end{eqnarray}

\item{\texttt{Minimal model 2:} $(2k_F+k_{10}\,, 2k_F+k_{\overline{126}}) = (10,8)$}
\\
For $k_{10}=0$, we have $k_F=5$ and $k_{\overline{126}}=-2$. Different from \texttt{minimal model 1}, the weight 8 modular forms of level 3 enter into the Yukawa couplings of $\overline{\Delta}$, and the superpotenital is given by
\begin{eqnarray}
\nonumber\mathcal{W}_Y &=&
\alpha_1 Y_{\mathbf{1}}^{(10)} \psi \psi H
+\alpha_2 Y_{\mathbf{1}'}^{(10)} \psi \psi H
+\alpha_3 Y_{\mathbf{3}I}^{(10)} \psi \psi H
+\alpha_4 Y_{\mathbf{3}II}^{(10)} \psi \psi H
+\alpha_5 Y_{\mathbf{3}III}^{(10)} \psi \psi H \\ \label{eq:SP_minimal_model_2}
&&+\gamma_1 Y_{\mathbf{1}}^{(8)} \psi \psi \overline{\Delta}
+\gamma_2 Y_{\mathbf{1}'}^{(8)} \psi \psi \overline{\Delta}
+\gamma_3 Y_{\mathbf{1}''}^{(8)} \psi \psi \overline{\Delta}+\gamma_4 Y_{\mathbf{3}I}^{(8)} \psi \psi \overline{\Delta}
 +\gamma_5 Y_{\mathbf{3}II}^{(8)} \psi \psi \overline{\Delta}\,.
\end{eqnarray}
\item{\texttt{Minimal model 3:} $(2k_F+k_{10}\,, 2k_F+k_{\overline{126}}) = (10,10)$}
\\
Both $H$ and $\overline{\Delta}$ couple with the weight 10 modular forms, and the modular weights of fields can be taken as $k_F=5$, $k_{10}=k_{\overline{126}}=0$. Thus we can read off the superpotential as follows,
\begin{eqnarray}
\nonumber\mathcal{W}_Y &=&
\alpha_1 Y_{\mathbf{1}}^{(10)} \psi \psi H
+\alpha_2 Y_{\mathbf{1}'}^{(10)} \psi \psi H
+\alpha_3 Y_{\mathbf{3}I}^{(10)} \psi \psi H
+\alpha_4 Y_{\mathbf{3}II}^{(10)} \psi \psi H+\alpha_5 Y_{\mathbf{3}III}^{(10)} \psi \psi H \\ \label{eq:SP_minimal_model_3}
&&+\gamma_1 Y_{\mathbf{1}}^{(10)} \psi \psi \overline{\Delta}
+\gamma_2 Y_{\mathbf{1}'}^{(10)} \psi \psi \overline{\Delta}
+\gamma_3 Y_{\mathbf{3}I}^{(10)} \psi \psi \overline{\Delta}+\gamma_4 Y_{\mathbf{3}II}^{(10)} \psi \psi \overline{\Delta}
+\gamma_5 Y_{\mathbf{3}III}^{(10)} \psi \psi \overline{\Delta}\,.
\end{eqnarray}

\end{itemize}
For each term of each model the explicit Yukawa matrices can be constructed in terms of
the complex coefficients $\alpha_a, \gamma_c$ and the
modular forms $Y_{\mathbf{r}}^{(k)}(\tau)$ as described in the previous subsection.
The total mass matrices will also depend on the mixing parameters $r_1,r_2$ and the VEVs
$v_u,v_d$ plus the neutrino parameters as in Eq.~\eqref{yuk}.
The above three minimal $SO(10)$ models with $A_4$ modular symmetry will be confronted with the experimental data, and the results are listed in table~\ref{tab:fit-minimal_model}. We see that the experimental data of quark and lepton masses and mixing parameters can be accommodated except the \texttt{minimal model 1}, the details would be discussed in section~\ref{section4}.

\subsection{Non-minimal models}

It is known that the minimal model is highly constrained in explaining the fermion masses and mixings. The Higgs field $\Sigma$ which is a  $\mathbf{120}$ dimensional multiplet of $SO(10)$ make it easy to account for the different masses and mixing patterns of the quarks and leptons, because the antisymmetric Yukawa coupling matrix $\mathcal{Y}^{120}$ have different coefficients in the Dirac mass matrices of the up-type quarks, down-type quarks, charged leptons and neutrinos, as shown in Eq.~\eqref{eq:mass-matrices-para}. As a consequence, the experimental data can be accommodated by using less terms than the minimal $SO(10)$ models. The superpotential is strongly constrained by modular symmetry and it is fixed by the modular weights of matter fields and Higgs fields. Since only the combination of the modular weights of matter fields and Higgs fields is relevant, the modular weight $k_{10}$ can be taken to be vanishing without loss of generality. For illustration, we present three typical models with small number of free parameters in the following.

\begin{itemize}[labelindent=-1.5em, leftmargin=1.6em]
\item{\texttt{Non-minimal model 1:} $(2k_F+k_{10}\,, 2k_F+k_{120}\,, 2k_F+k_{\overline{126}})=(4, 8, 0)$  }
\\
As an example, we can choose the values of modular weights as $k_F=2$, $k_{10}=0$, $k_{120}=4$ and $k_{\overline{126}}=-4$. The modular invariant superpotenital takes the following form
\begin{eqnarray} \label{eq:SP_non_minimal_model_1}
\mathcal{W}_Y &=&
\alpha_1 Y_{\mathbf{1}}^{(4)} \psi \psi H
+\alpha_2 Y_{\mathbf{1}'}^{(4)} \psi \psi H
+\alpha_3 Y_{\mathbf{3}}^{(4)} \psi \psi H
+\beta_1 Y_{\mathbf{3}I}^{(8)} \psi \psi \Sigma+\beta_2 Y_{\mathbf{3}II}^{(8)} \psi \psi \Sigma +
\gamma_1 \psi \psi \overline{\Delta}\,.
\end{eqnarray}
It is remarkable that this Yukawa superpotential is quite simple and it contains only six independent terms.

\item{\texttt{Non-minimal model 2:} $(2k_F+k_{10}\,, 2k_F+k_{120}\,, 2k_F+k_{\overline{126}})=(4,2,4)$  }
\\
The modular weights are determined to be $k_F=2$, $k_{120}=-2$, $k_{\overline{126}}=0$ for $k_{10}=0$. Modular invariance requires that the weight 2 and weight 4 modular forms of level 3 should be involved in the Yukawa coupling, and the superpotenital is given by
\begin{eqnarray}
\nonumber\mathcal{W}_Y &=&
\alpha_1 Y_{\mathbf{1}}^{(4)} \psi \psi H
+\alpha_2 Y_{\mathbf{1}'}^{(4)} \psi \psi H
+\alpha_3 Y_{\mathbf{3}}^{(4)} \psi \psi H
+\beta_1 Y_{\mathbf{3}}^{(2)} \psi \psi \Sigma \\ \label{eq:SP_non_minimal_model_2}
&&+\gamma_1 Y_{\mathbf{1}}^{(4)} \psi \psi \overline{\Delta}
+\gamma_2 Y_{\mathbf{1}'}^{(4)} \psi \psi \overline{\Delta}
+\gamma_3 Y_{\mathbf{3}}^{(4)} \psi \psi \overline{\Delta}\,.
\end{eqnarray}
\item{\texttt{Non-minimal model 3:} $(2k_F+k_{10}\,, 2k_F+k_{120}\,, 2k_F+k_{\overline{126}}) = (4,6,4)$ }
\\
This model differs from the \texttt{non-minimal model 2} in the modular weight of $\Sigma$, and the Yukawa superpotential reads as
\begin{eqnarray}
\nonumber\mathcal{W}_Y &=&
\alpha_1 Y_{\mathbf{1}}^{(4)} \psi \psi H
+\alpha_2 Y_{\mathbf{1}'}^{(4)} \psi \psi H
+\alpha_3 Y_{\mathbf{3}}^{(4)} \psi \psi H+\beta_1 Y_{\mathbf{3}I}^{(6)} \psi \psi\Sigma \\ \label{eq:SP_non_minimal_model_3}
&&+\beta_2 Y_{\mathbf{3}II}^{(6)} \psi \psi\Sigma +\gamma_1 Y_{\mathbf{1}}^{(4)} \psi \psi \overline{\Delta}
+\gamma_2 Y_{\mathbf{1}'}^{(4)} \psi \psi \overline{\Delta}
+\gamma_3 Y_{\mathbf{3}}^{(4)} \psi \psi \overline{\Delta} \,.
\end{eqnarray}
\end{itemize}
Similar to the previous minimal models, for each term of each model the explicit Yukawa matrices can be constructed in terms of
the complex coefficients $\alpha_a, \beta_b, \gamma_c$ and the
modular forms $Y_{\mathbf{r}}^{(k)}(\tau)$.
The total mass matrices will also depend on the mixing parameters $r_1,r_2,r_3$ and the VEVs
$v_u,v_d$ plus the neutrino parameters as in Eq.~\eqref{yuk}.
A comprehensive numerical analysis is performed for the above three models, as described in section~\ref{section4}. It can be seen that excellent agreement with experimental data can be achieved for certain values of the free parameters.

\section{Numerical analysis}
\label{section4}

A priori we cannot know which type of seesaw dominates or if they are of the same order of magnitude. In this section, we shall perform a detailed analysis of the predictions of the benchmark $SO(10)$ models given in section~\ref{section3}, and we shall consider the type I seesaw dominant scenario, type II seesaw dominant scenario and the mixture of both seesaw mechanisms in the neutrino sector. For any given values of the coupling constants and the complex modulus $\tau$, one can numerically diagonalize the fermion masses and then the mass eigenvalues, mixing angles and CP violation phases of both quarks and leptons can be extracted. The variation of the model parameters will dynamically affect the values of these experimental observables. In order to quantitatively estimate how well the above $SO(10)$ modular models can reproduce the fermion masses and mixing patterns at the GUT scale, we perform a $\chi^2$ analysis to find out the best fit values of the free parameters and the corresponding predictions for fermion masses and mixing parameters. Our full set of observables to which the models are fitted are the masses of quarks and charged leptons, the solar and atmospheric neutrino mass-squared differences, mixing angles of quarks and leptons, the CP violation phases $\delta_{CP}^{q}$ and $\delta_{CP}^{l}$ in the CKM matrix and lepton mixing matrix. The $\chi^{2}$ function is defined as
\begin{equation}
\chi^2=\sum_{i}\left(\frac{P_i(x)-\mu_i}{\sigma_i}\right)^2\,,
\end{equation}
where $P_i$ is the theoretical predictions as functions of the parameter set $x=\{\tau, \alpha_i, \beta_i,\gamma_i, r_1, r_2, r_3, c_{e}, \\ c_{\nu}, v_L, v_R\}$,  $\mu_i$ are the experimental values extrapolated to the GUT scale, and $\sigma_i$ denote the $1\sigma$ errors in $\mu_i$. Note that we have redefined the coupling constants $\alpha_i$, $\beta_i$ and $\gamma_i$ to absorb the coefficients $\frac{v^{10}_u}{v_u}$, $\frac{v^{\overline{126}}_d}{v^{10}_d}\frac{v^{10}_u}{v_u}$ and $\frac{v^{120}_d+v'^{120}_d}{v^{10}_d}\frac{v^{10}_u}{v_u}$ respectively. The $\chi^2$ sum runs over the seven mass ratios between charged fermions, ratio of the solar to atmospheric mass squared differences, the mixing angles and Dirac CP phases of quarks and leptons. We collect the values of $\mu_i$ and $\sigma_i$ underlying our analysis in table~\ref{Tab:parameter_values}, where the charged fermion masses and the quark mixing parameters measured at the electroweak scale have been evolved up to the corresponding ones at the GUT scale. The normal ordering neutrino masses are favored at $2.7\sigma$ level if the atmospheric neutrino data of Super-Kamiokande is taken into account~\cite{Esteban:2020cvm}. Therefore we assume normal ordering spectrum, the neutrino masses and mixing that we use are the low scale values taken from NuFIT 5.0~\cite{Esteban:2020cvm}. and we have ignored the effects of the evolution from the low energy scale to the GUT scale. The effects of evolution on the neutrino mass ratios and on the mixing angles are known to be negligible to a good approximation for the normal hierarchical spectrum.

In order to fit the model parameters to the observables, we numerically minimize $\chi^2$ with respect to free parameter vector $x$. We use the numerical minimization algorithm \texttt{TMinuit}~\cite{minuit} developed by CERN to numerically minimize the $\chi^2$ function to determine the best fit values of the input parameters. The mechanism of moduli stabilization is still an open question at present, consequently we treat the modulus $\tau$ as a free parameter to adjust the agreement with the data, and it randomly varies in the fundamental domain $\{\tau|\text{Re}\tau|\leq1/2, \text{Im}\tau>0, |\tau|\geq1\}$. After performing minimization of the $\chi^2$-function, we evaluate the fermion masses and mixing parameters with the best fit values of the free parameters. The overall scale $\alpha_1 v_u$ of the up quark mass matrix is fixed by the top quark mass. The down quarks and charged lepton matrices share the same factor $\alpha_1 r_1 v_d$ which is fixed by the measured value of the bottom quark mass. The scale factor of the light neutrino mass matrix is $\alpha_1 v^2_u/v_R$ for type I seesaw and $\alpha_1 v_L$ for type II seesaw and its value is determined by the solar neutrino mass splitting $\Delta m^2_{21}$.

Because the parameter space is high dimensional and the $\chi^2$ is a non-linear and complex function of the free parameters, generally many local minima exist. It is impossible to numerically determine whether a minimum is the global minimum of the $\chi^2$ function under consideration. We have repeated the minimization procedure many times with different initial values of the free parameters, and then choose the lowest one out of the many local minima found by the program. However, it is difficult to rule out the existence of still lower minima and predictions may be improved if they exist.

In order to quantitatively measure the degree of fine-tuning needed in the models, we use the fine-tuning factor $d_{FT}$ which was firstly introduced in~\cite{Altarelli:2010at}. The parameter $d_{FT}$ is defined as
\begin{equation}
\label{fine-tuning}
d_{FT}=\sum_{i} \Big| \frac{par_i}{err_i} \Big|\,,
\end{equation}
where $par_{i}$ is the best fit value of $i^{\text{th}}$ real input parameter and $err_i$ is the corresponding offset of $par_{i}$ which changes $\chi^2$ by one unit with all other parameters fixed at their best fit values. Notice that $err_i$ is not the uncertainty of the experimental data given in table~\ref{Tab:parameter_values}. If some $|err_i/par_i|$ are very small, then a small variation of the corresponding parameters would make a large difference on the $\chi^2$. Hence $d_{FT}$ can roughly measure the amount of fine-tuning involved in the fit. Similar to Ref.~\cite{Altarelli:2010at}, one can understand the significance of the fine-tuning by comparing $d_{FT}$ with a similar parameter $d_{Data}$ derived from the data:
\begin{equation}
\label{data-precision}
d_{Data}=\sum_{i}\Big| \frac{\mu_{i}}{\sigma_i}\Big|\,,
\end{equation}
where $\mu_i$ and $\sigma_i$ are the central values and the $1\sigma$ of the  $i^{\text{th}}$ observables respectively. We have $d_{Data}=531.986$ for the set of the experimental data in table~\ref{Tab:parameter_values}.

As shown in Eq.~\eqref{eq:mnu-I-II-seesaw}, generally both type I and type II seesaw can contribute to the light neutrino masses in $SO(10)$ GUTs. A priori, we do not know which type of seesaw dominates or if they are of the same order of magnitude. In the following, we shall perform a detailed analysis of the predictions of the benchmark $SO(10)$ models given in section~\ref{section3}, and we shall consider three scenarios: the type I seesaw dominance, the type II seesaw dominance and the mixture of both seesaw mechanisms in the neutrino sector.

\subsection{Numerical results of the minimal models
}

In this section, we consider the numerical results of the models with the minimal Higgs content $H$ and $\overline{\Delta}$ which are in the $SO(10)$ representations $\mathbf{10}$ and $\mathbf{\overline{126}}$ respectively. As a result, the term $\mathcal{\widetilde{Y}}^{120}$ is absent in the fermion mass matrices of Eq.~\eqref{eq:mass-matrices-para}. From Eqs.~(\ref{eq:Y_10_126}, \ref{eq:Yukawa_Y_120}) we see that the forms of Yukawa couplings are determined by the modular weights $2k_{F}+k_{10}$ and $2k_{F}+k_{\overline{126}}$, and they depend on a number of coupling constants and the complex modulus $\tau$. In this work, we will be concerned with the modular forms of level 3 up to weight 10, and higher weight modular forms can be discussed in a similar way although more modular invariant contractions as well as couplings would be involved.
Thus the possible values of $2k_{F}+k_{10}$ and $2k_{F}+k_{\overline{126}}$ are 0, 2, 4, 6, 8, 10 up to weight 10 modular forms, and consequently we can obtain $6\times 6=36$ minimal $SO(10)$ GUT models.

We numerically analyze all these models and minimize the corresponding $\chi^{2}$ function with the TMinuit package~\cite{minuit}. We find that only two out of the 36 models can give a good fit to the experimental data for certain values of input parameters. Nevertheless some minimal models with less parameters can achieve a relatively small $\chi^{2}_{\text{min}}$, and only the quark mass ratio $m_{d}/m_{s}$ lies outside the $3\sigma$ allowed region. It's reasonable to regard these models as a good leading order approximation, we would like to give such an example model named as $\texttt{minimal model 1}$. This model is specified by the sum of modular weights $(2k_{F}+k_{10}, 2k_{F}+k_{\overline{126}})=(10,6)$ and the corresponding superpotential is given in Eq.~\eqref{eq:SP_minimal_model_1}.
We find that if both types of seesaw are present, almost all fermion observables can be correctly reproduced except the $m_{d}/m_{s}$ ratio. The fitting results of the $\texttt{minimal model 1}$ is summarized in table~~\ref{tab:fit-minimal_model}, where we split the $\chi^{2}$ into three parts $\chi^{2}=\chi^{2}_{l}+\chi^{2}_{q}+\chi^{2}_{b\tau}$ and $\chi^2_l$, $\chi^2_q$ and $\chi^2_{b\tau}$ denote the pieces of $\chi^2$ function arising from the deviations of the lepton sector observables, the quark sector observables and $m_b/m_{\tau}$ from their central values respectively. We see that $m_{d}/m_{s}$ is predicted to be small and it is about $5\sigma$ away from the experimental best fit value
 while all other observables are within the $3\sigma$ ranges of the experimental data.

The good agreement with data can be achieved with higher modular weight $2k_{F}+k_{\overline{126}}$ but at the price of introducing more free parameters in the model. The two phenomenologically viable minimal models are named as $\texttt{minimal model 2}$ and $\texttt{minimal model 3}$ with modular weights $(2k_{F}+k_{10}, 2k_{F}+k_{\overline{126}})$ equal to $(10, 8)$ and $(10, 10)$ respectively. The corresponding forms of the Yukawa superpotentials are given in Eq.~\eqref{eq:SP_minimal_model_2} and Eq.~\eqref{eq:SP_minimal_model_3}. Comprehensive numerical analysis shows that $\texttt{minimal model 2}$ and $\texttt{minimal model 3}$ can accommodate the experimental data if both type-I and type-II seesaw mechanisms contribute to the light neutrino masses. There are total $27$ free real parameters in fermion mass matrices in both models. Our fits yield $\chi^{2}=8.59476$ for the $\texttt{minimal model 2}$ and $\chi^{2}=3.60588$  for the $\texttt{minimal model 3}$. We tabulate the best fit values of the input parameters and the predictions for the fermion observables in table~\ref{tab:fit-minimal_model}.
It is notable that almost all the measured observables fall within the $1\sigma$ experimentally allowed ranges for these two viable $SO(10)$ minimal models. Moreover, we give the model predictions for several as yet unmeasured observables including the Majorana CP violation phases $\alpha_{21}$, $\alpha_{31}$, the lightest neutrino mass $m_{1}$, the masses of the heavy right-handed neutrinos $M_{1,2,3}$ and the effective mass $m_{\beta\beta}$ in
neutrinoless double beta decay, they can be understood as predictions of the models. The effective Majorana neutrino mass $m_{\beta\beta}$ are defined as
\begin{equation}
m_{\beta\beta}=\left| m_1\cos^2\theta^{l}_{12}\cos^2\theta^l_{13}+m_2
\sin^2\theta^{l}_{12}\cos^2\theta^l_{13}e^{i\alpha_{21}}+m_3\sin^2\theta^l_{13}e^{i(\alpha_{31}-2\delta^{l}_{CP})}\right|\,.
\end{equation}
The predicted value of $m_{\beta\beta}$ at the best-fit point in the $\texttt{minimal model 2}$ and $\texttt{minimal model 3}$ are $m_{\beta\beta}=0.738522~\text{meV}$ and $m_{\beta\beta}=0.667982~\text{meV}$ respectively, which are far below the sensitivity of future tonne-scale neutrinoless double beta decay experiments. We can use the measured value of the solar neutrino mass squared $\Delta m_{21}^{2}=7.42\times 10^{-5}~\text{eV}^{2}$~\cite{Esteban:2020cvm} to fix the overall mass scale $\alpha_1v^2_u/v_R$ of neutrino mass matrix, subsequently the absolute values of light neutrino masses can be determined as follows,
\begin{eqnarray} \nonumber
&&\texttt{Minimal model 2:}\quad m_{1}=4.91455~\text{meV}, \quad
     m_{2}=9.91730~\text{meV},\quad
     m_{3}=50.43670~\text{meV}\,,\\
&&\texttt{Minimal model 3:}\quad m_{1}=5.14187~\text{meV}, \quad
     m_{2}=10.03190~\text{meV},\quad
     m_{3}=50.54390~\text{meV}\,.
\end{eqnarray}
The seesaw scale is of order $10^{10}\sim 10^{11}~\text{GeV}$ in the two models.

\begin{figure}[hptb!]
\centering
\includegraphics[width=6.5in]{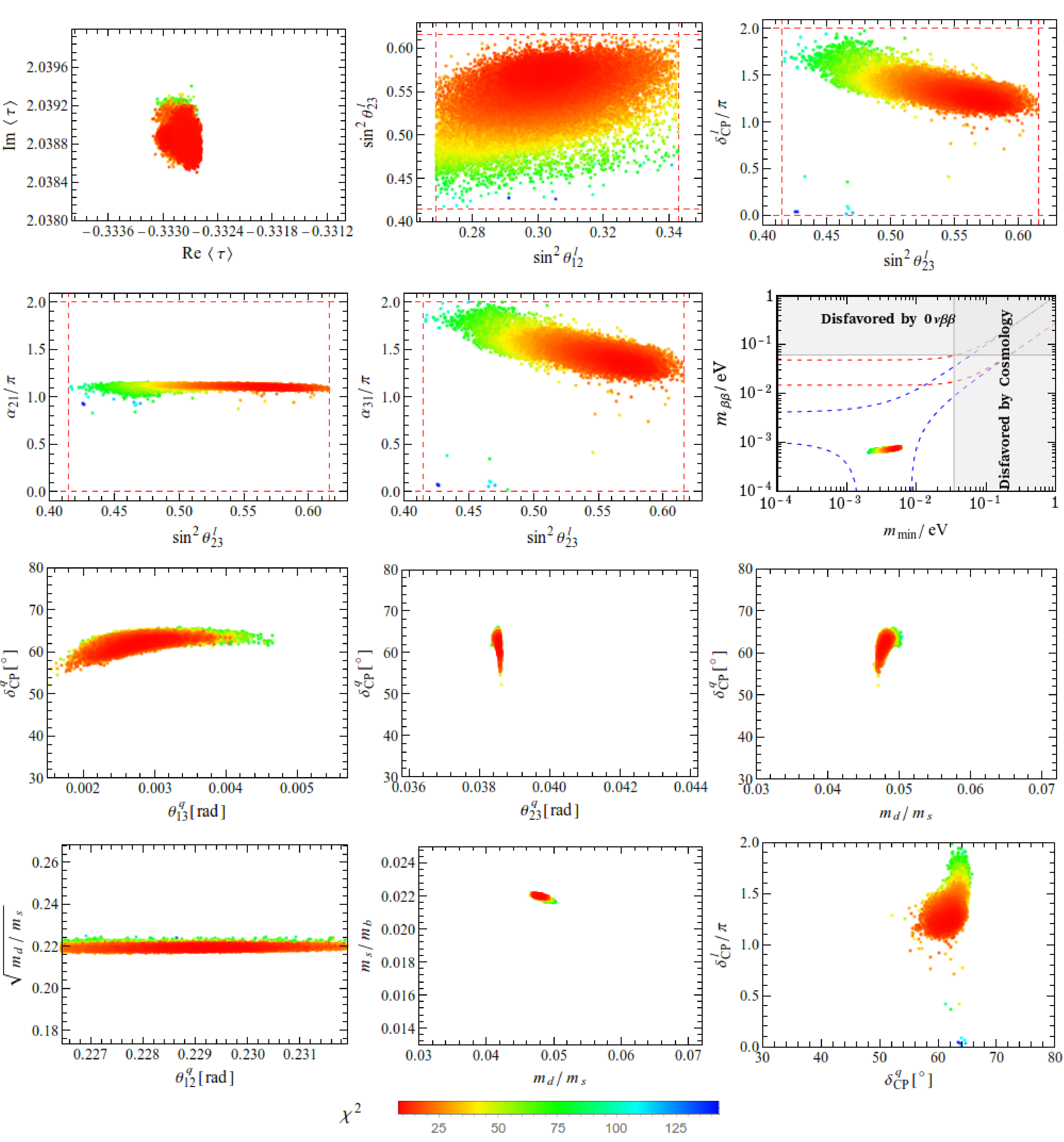}
\caption{
The values of the complex modulus $\tau$ compatible with experimental data and the correlations between the neutrino mixing angles, CP violation phases, quark mass ratios and mixing parameters in the $\texttt{minimal model 2}$ with $\texttt{SS-I+II}$. The vertical and horizontal dashed lines are the $3\sigma$ bounds taken from~\cite{Esteban:2020cvm}. In the panel of $m_{\beta\beta}$ with respect to the lightest neutrino mass $m_{\text{min}}$, the blue (red) dashed lines stand for the most general allowed regions for normal ordering (inverted ordering) neutrino mass spectrum respectively, where the neutrino oscillation parameter are varied within their $3\sigma$ ranges. The present upper limit $m_{\beta\beta}<(61-165)$ meV from KamLAND-Zen~\cite{KamLAND-Zen:2016pfg} is shown by horizontal grey band. The vertical grey exclusion band represents the most aggressive bound $\sum_im_i<0.120$eV from Planck~\cite{Planck:2018vyg}. }
\label{fig:minimal_model_1}
\end{figure}

\begin{figure}[hptb!]
\centering
\includegraphics[width=6.5in]{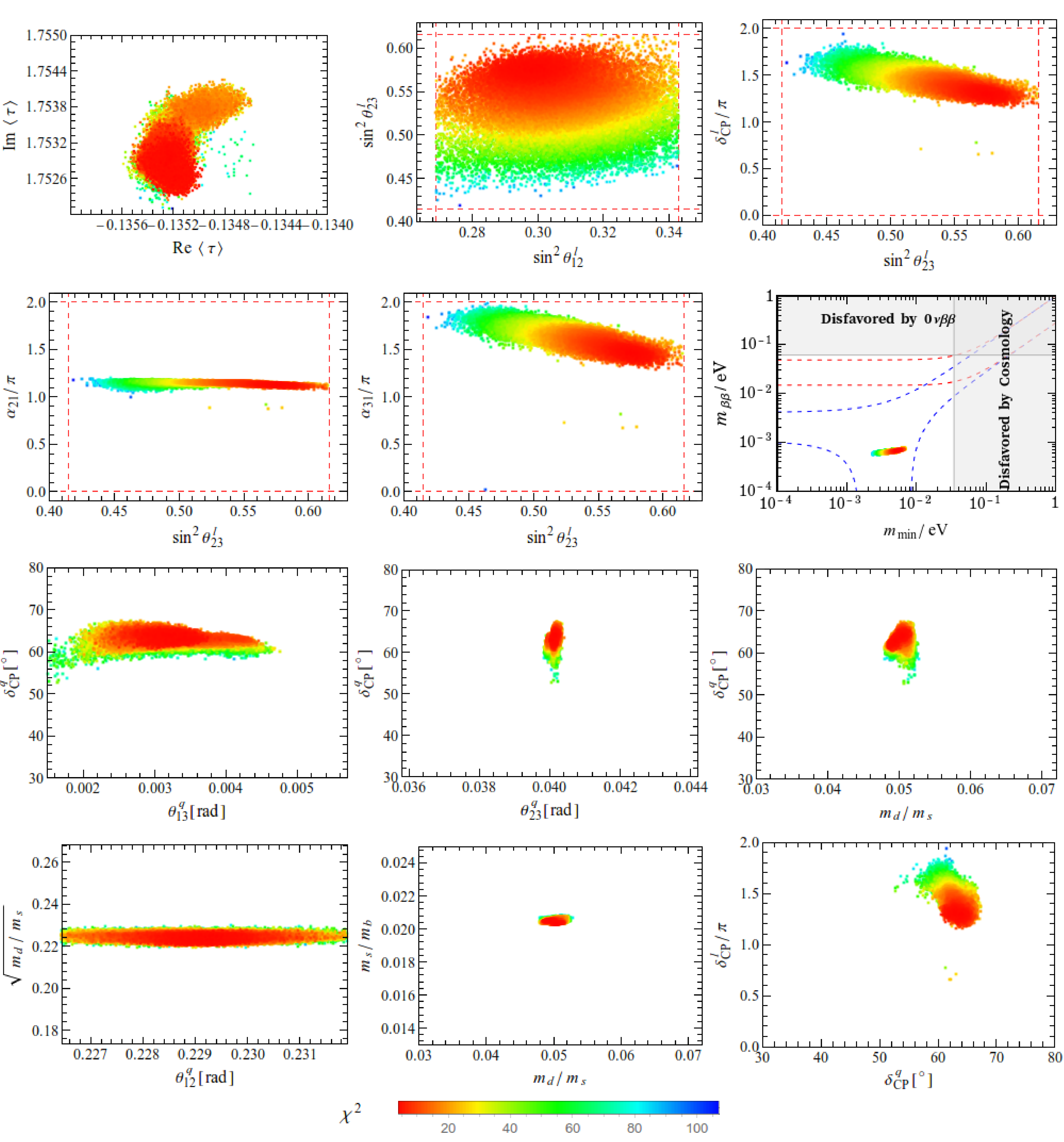}
\caption{The values of the complex modulus $\tau$ compatible with experimental data and the correlations between the neutrino mixing angles, CP violation phases, quark mass ratios and mixing parameters in the $\texttt{minimal model 3}$ with $\texttt{SS-I+II}$. The vertical and horizontal dashed lines are the $3\sigma$ bounds taken from~\cite{Esteban:2020cvm}.}
\label{fig:minimal_model_2}
\end{figure}

Furthermore, we use the widely-used sampling program  \texttt{MultiNest}~\cite{Feroz:2007kg,Feroz:2008xx} to scan the parameter space around the best fit points, and the predictions for the fermion masses and mixing parameters are required to be compatible with data at $3\sigma$ level. Notice that different predictions could possibly be obtained around other local minima. The allowed values of the modulus $\tau$ and the correlations between observables for the $\texttt{minimal model 2}$ and $\texttt{minimal model 3}$ are plotted in figure~\ref{fig:minimal_model_1} and figure~\ref{fig:minimal_model_2} respectively. We can see that the regions of $\tau$ compatible with data are very narrow in the two minimal models, the effective Majorana mass $m_{\beta\beta}$ characterizing the neutrinoless double beta decay amplitude lies in the narrow intervals around $0.7~\text{meV}$ which is too small to be detected.

Although both $\texttt{minimal model 2}$ and $\texttt{minimal model 3}$ can give good accommodation to the experimental results,  we note that a substantial amount of fine tuning of the free parameters is needed. The fine-tuning parameter $d_{FT}$ defined in Eq.~\eqref{fine-tuning} is found to be of order $d_{FT}\approx 10^{6}$ in both $\texttt{minimal model 2}$ and $\texttt{minimal model 3}$, to be compared with $d_{Data}=531.986$.

\subsection{Numerical results of the non-minimal models
}
In this section, we proceed to consider the numerical results of the non-minimal models with the full Higgs content $H$, $\overline{\Delta}$ and $\Sigma$ which are 10, 126 and 120 dimensional multiplets of $SO(10)$. For the non-minimal $SO(10)$ models, the general form of the Yukawa matrices for different fermions and the right-handed neutrino mass matrix are given in Eq.~\eqref{eq:mass-matrices-para}. In comparison with the minimal $SO(10)$ models, three more Higgs mixing parameters $r_3$, $c_e$ and $c_{\nu}$ accompanying the Yukawa coupling $\mathcal{Y}^{120}$ of $\Sigma$ are present. The $SO(10)$ non-minimal models are completely specified by the modular weights $2k_F+k_{10}$, $2k_F+k_{120}$ and $2k_F+k_{\overline{126}}$. Hence there are total $6\times6\times5=180$ possible non-minimal models if one concerns with the modular forms up to weight 10.

We scan the parameter space of the above non-minimal models one by one. In contrast with the minimal models, there are plenty of non-minimal models which are compatible with the experimental data of fermion masses and mixings even in either type-I or type-II dominated seesaw mechanisms. It is too lengthy to list all the viable models here. In the following, we present three benchmark non-minimal models which are characterized by
\begin{eqnarray}
&&\texttt{Non-minimal model 1:}~ (2k_F+k_{10}\,, 2k_F+k_{120}\,, 2k_F+k_{\overline{126}})=(4, 8, 0)\,,\\
&&\texttt{Non-minimal model 2:}~ (2k_F+k_{10}\,, 2k_F+k_{120}\,, 2k_F+k_{\overline{126}})=(4, 2, 4)\,,\\
&&\texttt{Non-minimal model 3:}~ (2k_F+k_{10}\,, 2k_F+k_{120}\,, 2k_F+k_{\overline{126}})=(4, 6, 4)\,.
\end{eqnarray}
The superpotentials can be straightforwardly read out as given in Eq.~\eqref{eq:SP_non_minimal_model_1}, Eq.~\eqref{eq:SP_non_minimal_model_2} and Eq.~\eqref{eq:SP_non_minimal_model_3} respectively. We report the fitting results of the above three non-minimal models in table~\ref{tab:fit-non_minimal_model_1}, table~\ref{tab:fit-non_minimal_model_2} and table~\ref{tab:fit-non_minimal_model_3} respectively.

\begin{figure}[hptb!]
\centering
\includegraphics[width=6.5in]{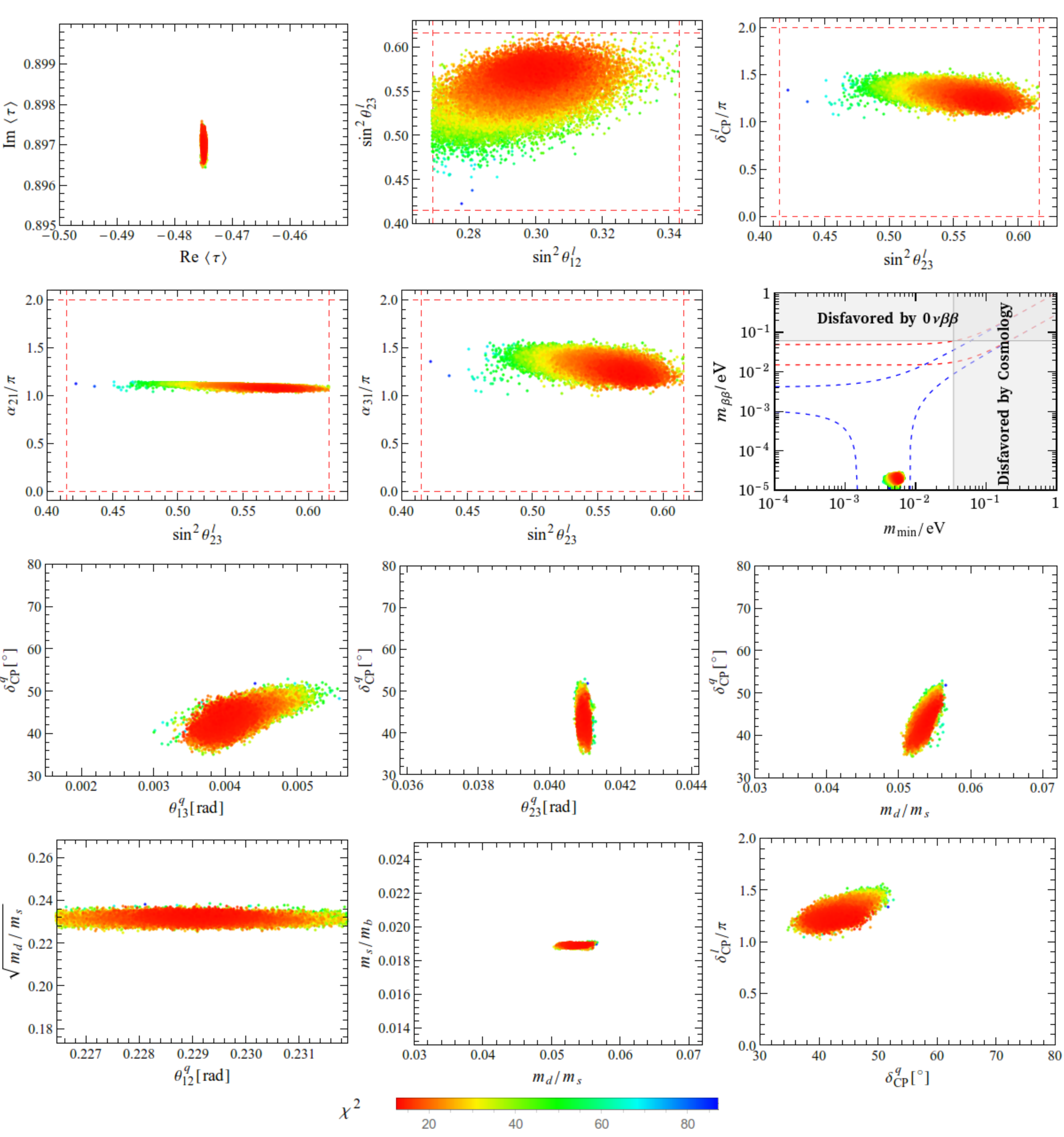}
\caption{The values of the complex modulus $\tau$ compatible with experimental data and the correlations between the neutrino mixing angles, CP violation phases, quark mass ratios and mixing parameters in the $\texttt{non-minimal model 1}$ with $\texttt{SS-I}$. The vertical and horizontal dashed lines are the $3\sigma$ bounds taken from~\cite{Esteban:2020cvm}. }
\label{fig:non_minimal_1_I}
\end{figure}

\begin{figure}[hptb!]
\centering
\includegraphics[width=6.5in]{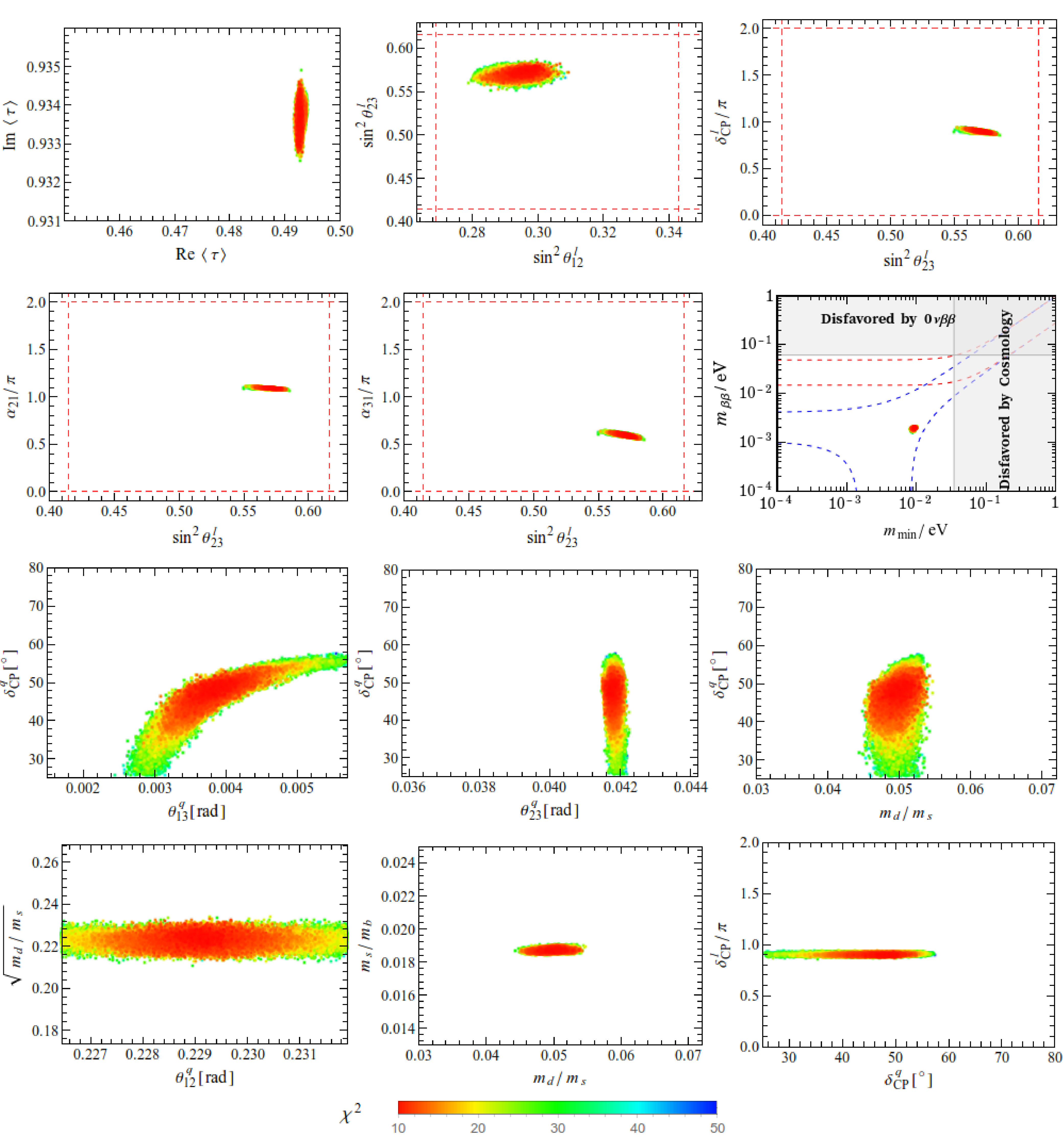}
\caption{The values of the complex modulus $\tau$ compatible with experimental data and the correlations between the neutrino mixing angles, CP violation phases, quark mass ratios and mixing parameters in the $\texttt{non-minimal model 1}$ with $\texttt{SS-I+II}$. The vertical and horizontal dashed lines are the $3\sigma$ bounds taken from~\cite{Esteban:2020cvm}.}
\label{fig:non_minimal_1_III}
\end{figure}

The $\texttt{non-minimal model 1}$ can give a reasonably good fit to all observables if the neutrino masses are generated by the type-I seesaw and the mixture of type-I and type-II seesaw which are denoted as $\texttt{SS-I}$ and $\texttt{SS-I+II}$ respectively. It turns out that this model is unable to reproduce the fermion masses and mixing for the pure type-II seesaw case, and accordingly the minima of $\chi^{2}$ from the numerical optimization algorithm is quite large. In the type-I seesaw dominant scenario, 23 real free parameters are involved and the main contribution to $\chi^2_{\rm min}$ comes from the mass ratio $m_b/m_{\tau}$ which is about $3\sigma$ larger than its central value, and all other observables are reproduced correctly with small deviations as shown in the table~\ref{tab:fit-non_minimal_model_1}. The light neutrino masses for the type-I and type-I+II cases are predicted to be
\begin{eqnarray}\nonumber
&&\texttt{SS-I:}~m_{1}=5.52462~\text{meV}, \quad
m_{2}=10.2333~\text{meV},\quad m_{3}=50.5706~\text{meV}\,,\\
&&\texttt{SS-I+II:}~ m_{1}=9.1019~\text{meV}, \quad m_{2}=12.5317~\text{meV},\quad m_{3}=51.0373~\text{meV}\,,
\end{eqnarray}
for the $\texttt{non-minimal model 1}$. The allowed values of $\tau$ and the correlations between observables are displayed in figure~\ref{fig:non_minimal_1_I}. The combination of type-I and type-II seesaw mechanism can give a better fit of the data but at the price of introducing one more complex free parameter $v_L$. The smallest $\chi^{2}$ is found to be $\chi^{2}_{\text{min}}=9.98822$ and among all observables the largest contributions to $\chi^{2}$ is $\chi^{2}_{b\tau}=3.10314$ from $m_b/m_{\tau}$. The corresponding correlations between predictions are plotted in figure~\ref{fig:non_minimal_1_III}. We can see from figure~\ref{fig:non_minimal_1_III} that the allowed regions of lepton mixing angles $\theta_{12}^{l}$, $\theta_{23}^{l}$ and CP-violation phases $\delta_{CP}^{l}$, $\alpha_{21}$ as well as $\alpha_{31}$ are quite narrow. The effective Majorana mass $m_{\beta\beta}$ is predicted to be around $1.9~\text{meV}$ and it is far below the sensitivities of next generation neutrinoless double beta decay experiments.

\begin{figure}[hptb!]
\centering
\includegraphics[width=6.5in]{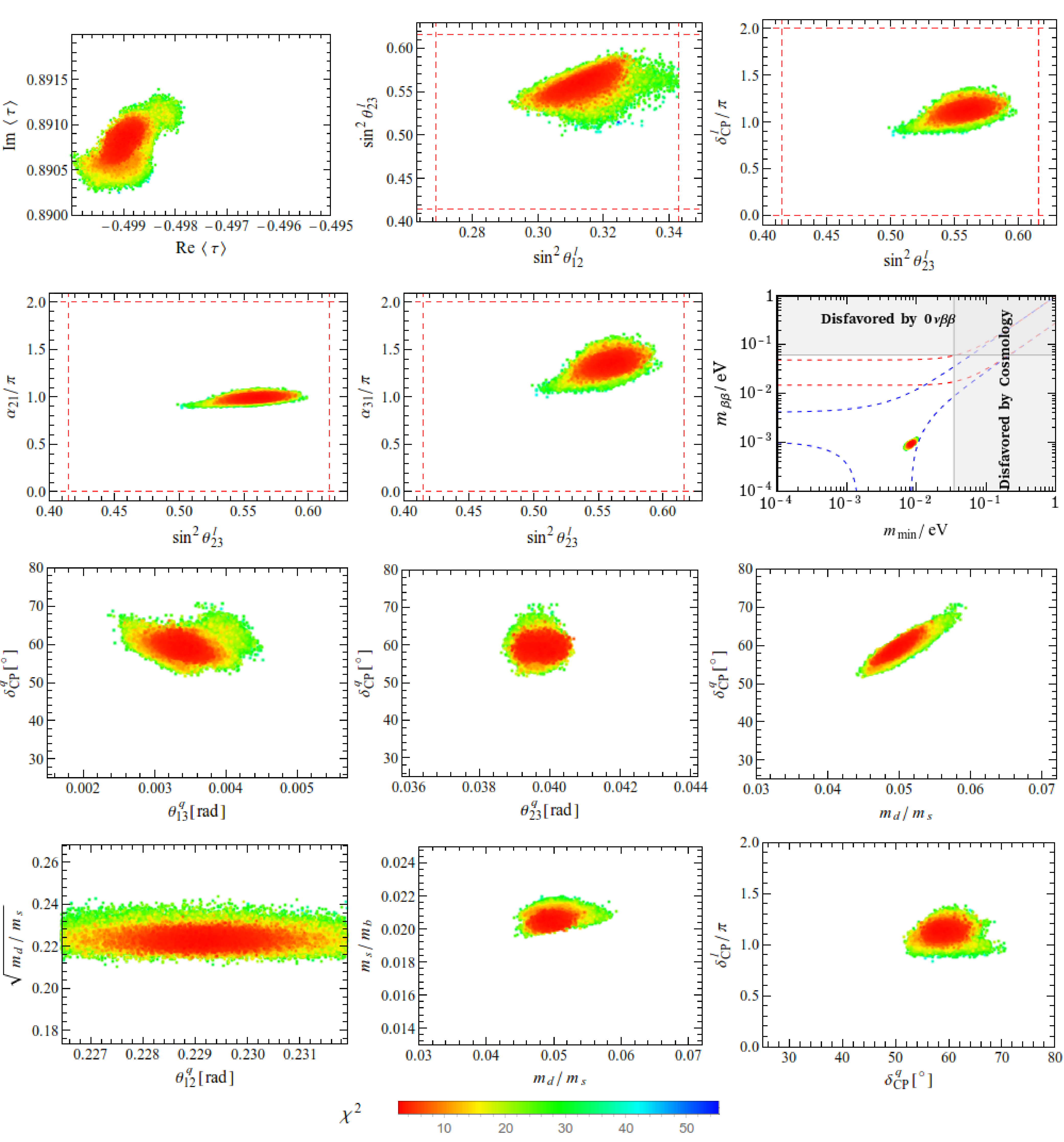}
\caption{The values of the complex modulus $\tau$ compatible with experimental data and the correlations between the neutrino mixing angles, CP violation phases, quark mass ratios and mixing parameters in the $\texttt{non-minimal model 2}$ with $\texttt{SS-I}$. The vertical and horizontal dashed lines are the $3\sigma$ bounds taken from~\cite{Esteban:2020cvm}.}
\label{fig:non_minimal_2_I}
\end{figure}

\begin{figure}[hptb!]
\centering
\includegraphics[width=6.5in]{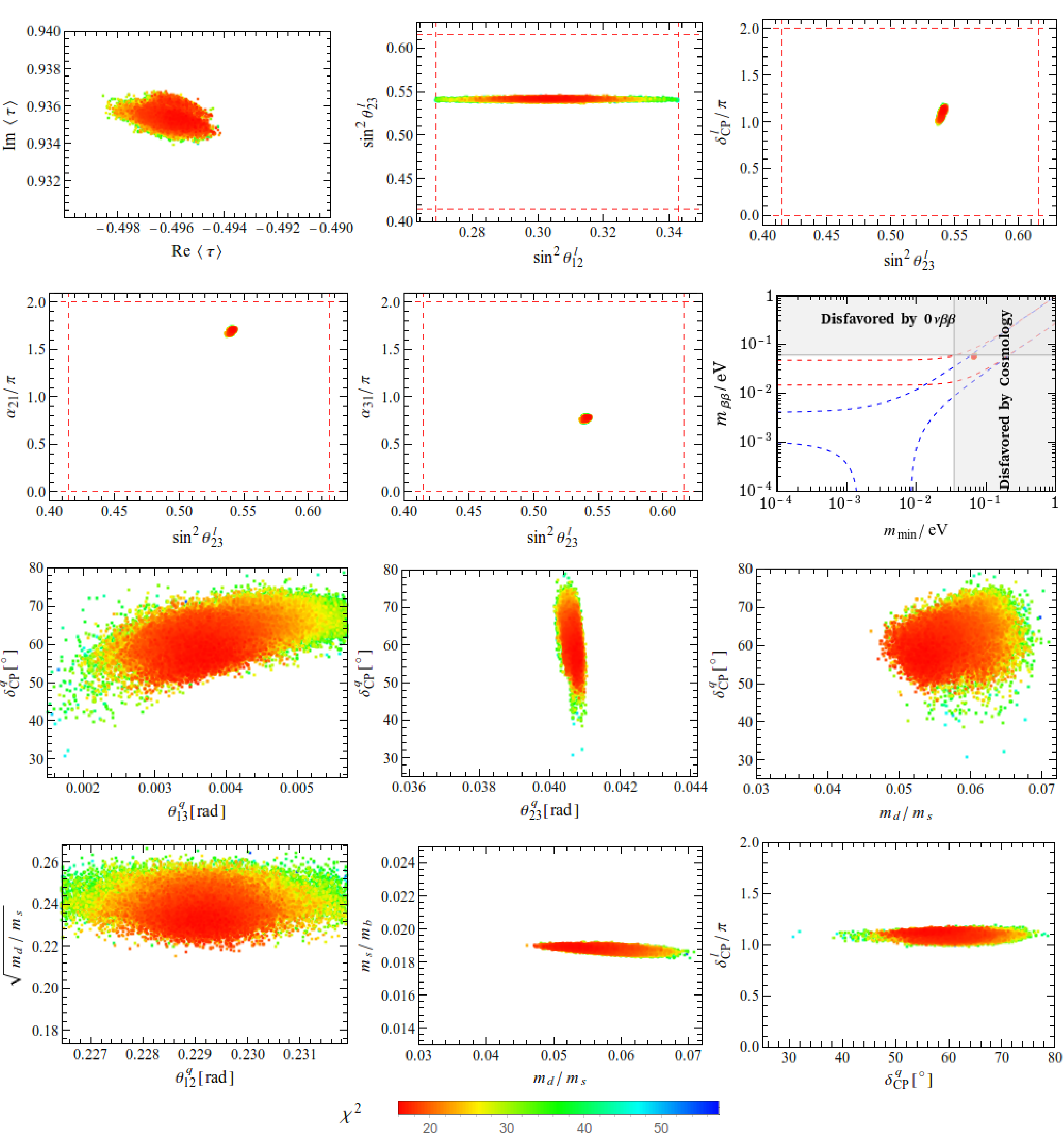}
\caption{The values of the complex modulus $\tau$ compatible with experimental data and the correlations between the neutrino mixing angles, CP violation phases, quark mass ratios and mixing parameters in the $\texttt{non-minimal model 2}$ with $\texttt{SS-II}$. The vertical and horizontal dashed lines are the $3\sigma$ bounds taken from~\cite{Esteban:2020cvm}. }
\label{fig:non_minimal_2_II}
\end{figure}

\begin{figure}[hptb!]
\centering
\includegraphics[width=6.5in]{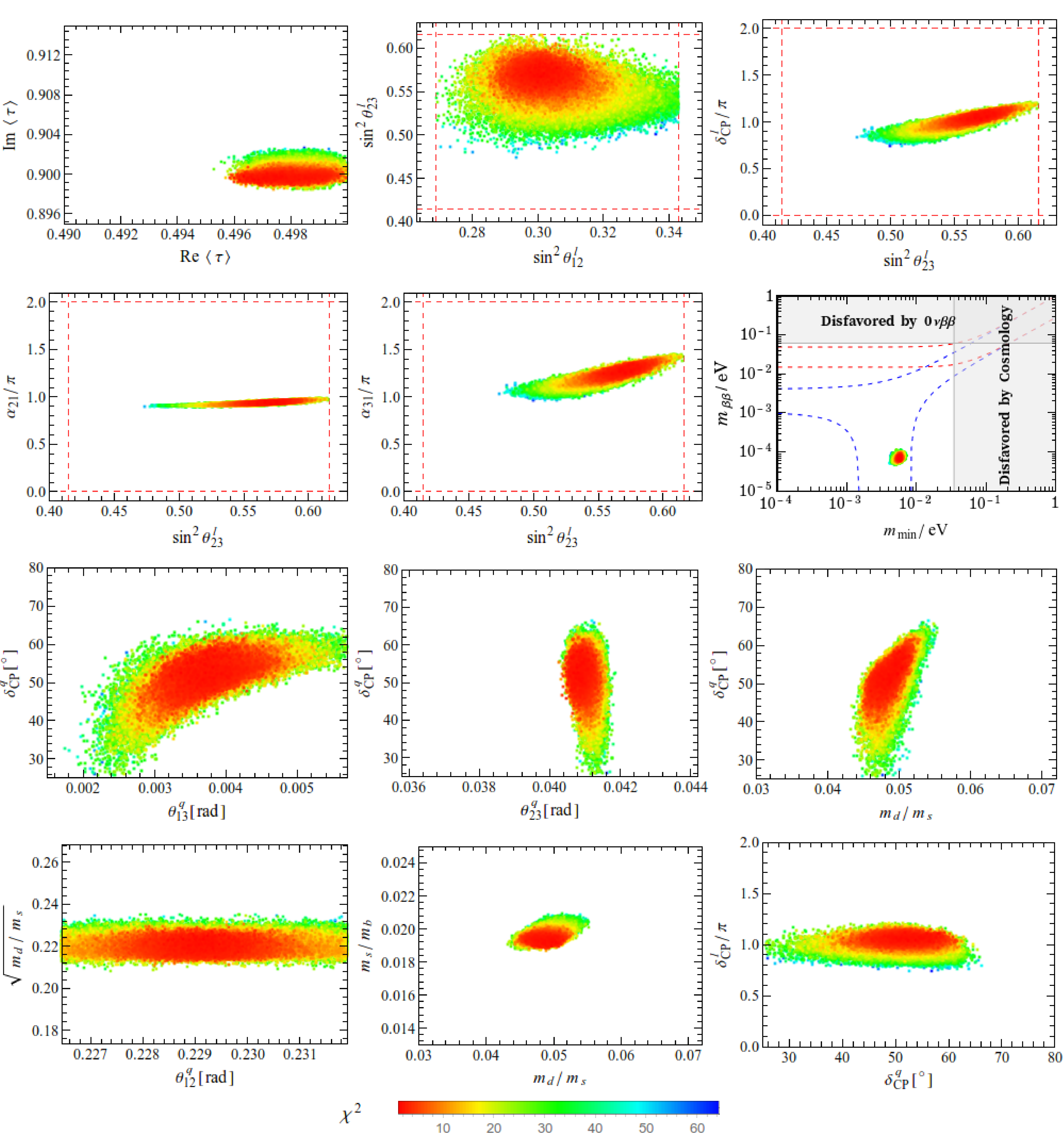}
\caption{The values of the complex modulus $\tau$ compatible with experimental data and the correlations between the neutrino mixing angles, CP violation phases, quark mass ratios and mixing parameters in the $\texttt{non-minimal model 2}$ with $\texttt{SS-I+II}$. The vertical and horizontal dashed lines are the $3\sigma$ bounds taken from~\cite{Esteban:2020cvm}.  }
\label{fig:non_minimal_2_III}
\end{figure}

The numerical fitting results of the $\texttt{non-minimal model 2}$ are summarized in table~\ref{tab:fit-non_minimal_model_2}, and this model can match the measured values of observables given in table~\ref{Tab:parameter_values} for all the three types of neutrino mass generation mechanism. The mass ratio $m_b/m_{\tau}$ is about $3.5\sigma$ away from its mean value in the type-II seesaw dominant scenario, while all other observables can be explained well with small derivations. The issue about the prediction of $m_{b}/m_{\tau}$ can be well resolved if the type-I seesaw mechanism dominates in neutrino sector, and the minimum of $\chi^{2}$ is found to be $\chi^{2}_{\text{min}}=2.71438$ with $\chi^{2}_{b\tau}=0.32689$. The best fits values of fermion masses and the mixing parameters are in good agreement with the corresponding inputted reference values. The fitting results can be slightly improved in the mixture of type-I and type-II seesaw mechanisms. The overall scale of the neutrino masses is fixed by the solar neutrino mass difference $\Delta m_{21}^{2}=7.42\times 10^{-5}~\text{eV}^{2}$, and the light neutrino masses are determined to be
\begin{eqnarray}\nonumber
&&\texttt{SS-I:}~ m_{1}=8.39159~\text{meV}, \quad m_{2}=12.0258~\text{meV},\quad m_{3}=50.7452~\text{meV}\,,\\ \nonumber
&&\texttt{SS-II:}~ m_{1}=66.9075~\text{meV}, \quad m_{2}=67.4597~\text{meV},\quad m_{3}=83.6365~\text{meV}\,,\\
&&\texttt{SS-I+II:}~ m_{1}=5.74271~\text{meV}, \quad m_{2}=10.3527~\text{meV},\quad m_{3}=50.6365~\text{meV}\,.
\end{eqnarray}
Notice that the neutrino mass spectrum is strongly hierarchical in the case of type-I and type-I+II seesaw, while it is quasi-degenerate in the pure type-II seesaw case. For the pure type-II seesaw case, the light neutrino masses are quasi-degenerate and the neutrino mass sum is $\sum_im_i=218.004$ meV which is above the aggressive upper bound 120meV but still below the conservative upper limit 600 meV given by the Planck collaboration~\cite{Planck:2018vyg}. Notice that the cosmological bound on the neutrino masses significantly depend on the data sets which are combined to break the degeneracies of the many cosmological parameters, and the upper limit on neutrino mass becomes weaker when one departs from the framework of $\Lambda$CDM
plus neutrino mass to frameworks with more cosmological parameters. Three heavy right-handed neutrinos are introduced in the type-I seesaw mechanism, and the scale of the right-handed neutrino mass are predicted to be $10^{10}\sim 10^{13}~\text{GeV}$ in the type-I and type-I+II cases. We show the allowed values of $\tau$ and the correlations between observables for \texttt{SS-I},  \texttt{SS-II} and  \texttt{SS-I+II} in figure~\ref{fig:non_minimal_2_I}, figure~\ref{fig:non_minimal_2_II} and figure~\ref{fig:non_minimal_2_III} respectively. Notice that the possible values of certain observables lie in rather small regions.

The $\texttt{non-minimal model 3}$ differs from the $\texttt{non-minimal model 2}$ in the value of $2k_{F}+k_{120}$. From Eq.~\eqref{eq:Yukawa_Y_120} we know that there are two independent $A_4$ invariant terms in the antisymmetric Yukawa coupling $\mathcal{Y}^{120}$ for $2k_{F}+k_{120}=6$ in the $\texttt{non-minimal model 3}$ while only a single term is allowed for $2k_{F}+k_{120}=2$ as in $\texttt{non-minimal model 2}$. As a consequence, the $\texttt{non-minimal model 3}$ will have one more complex coupling than the $\texttt{non-minimal model 2}$. The fitting results are given in table~\ref{tab:fit-non_minimal_model_3}, we see that all three types of seesaw mechanisms can give quite good fits to the fermion observables. The minimal $\chi^{2}_{\text{min}}=5.61610$ in pure type-II case is smaller than the $\chi^{2}_{\text{min}}=7.02930$ in the type-I seesaw dominant case. Therefore the type-II seesaw can explain the data a bit better than the type-I seesaw.
The predictions of light neutrino masses for three seesaw cases are
\begin{eqnarray}
\nonumber &&\texttt{SS-I:}~ m_{1}=6.33620~\text{meV}, \quad m_{2}=10.69330~\text{meV},\quad m_{3}=50.41520~\text{meV}\,,\\ \nonumber
&&\texttt{SS-II:}~ m_{1}=50.5718~\text{meV}, \quad m_{2}=51.3002~\text{meV},\quad m_{3}=71.2635~\text{meV}\,,\\
&&\texttt{SS-I+II:}~ m_{1}=7.104~\text{meV}, \quad m_{2}=11.1654~\text{meV},\quad m_{3}=50.6344~\text{meV}\,.
\end{eqnarray}
Analogous to the $\texttt{non-minimal model 2}$, the light neutrino masses are also quasi-degenerate for type-II seesaw dominance and they are consistent with the conservative limit on neutrino masses from Planck~\cite{Planck:2018vyg}.

\begin{figure}[hptb!]
\centering
\includegraphics[width=6.5in]{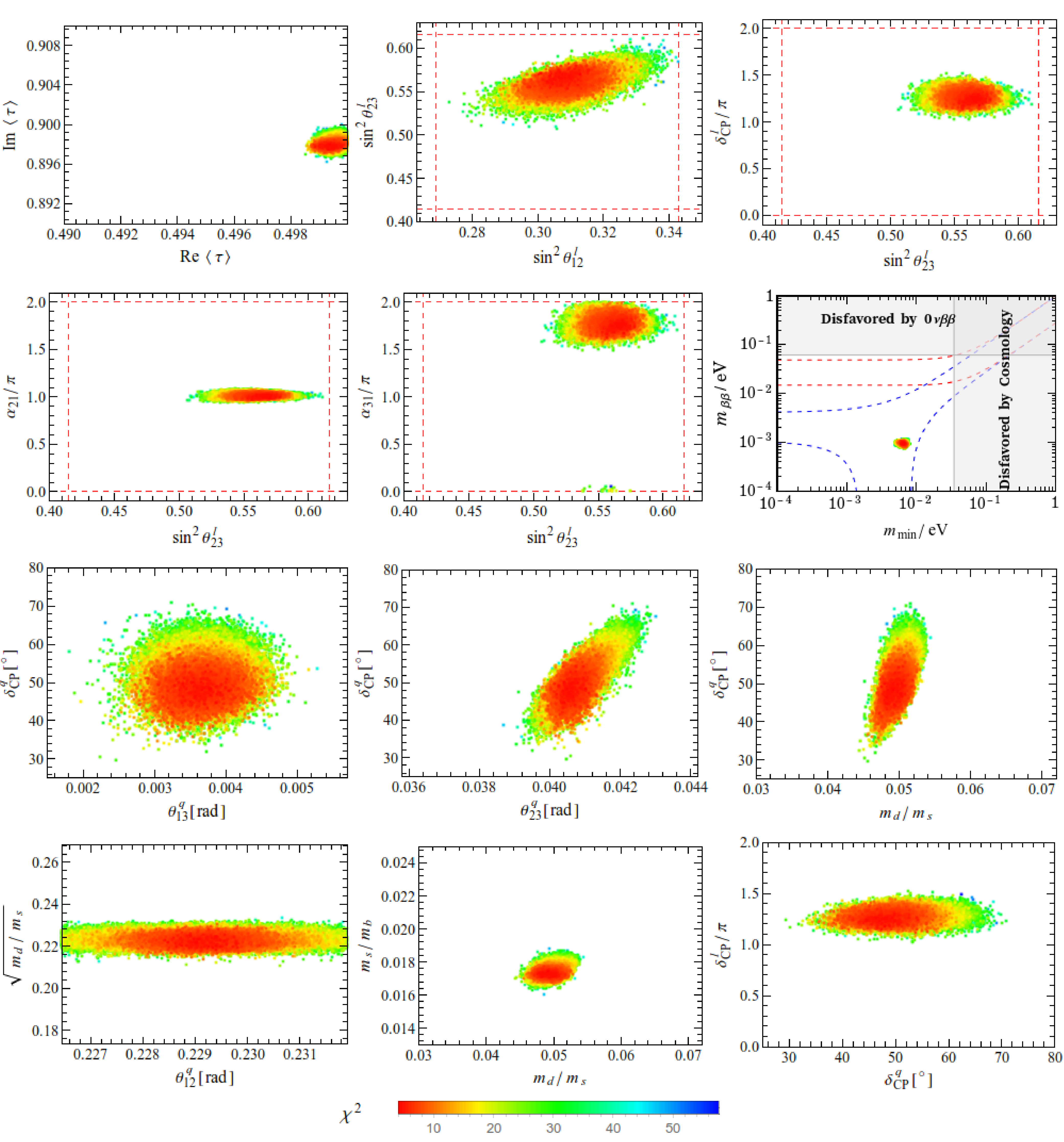}
\caption{The values of the complex modulus $\tau$ compatible with experimental data and the correlations between the neutrino mixing angles, CP violation phases, quark mass ratios and mixing parameters in the $\texttt{non-minimal model 3}$ with $\texttt{SS-I}$. The vertical and horizontal dashed lines are the $3\sigma$ bounds taken from~\cite{Esteban:2020cvm}.}
\label{fig:non_minimal_3_I}
\end{figure}

\begin{figure}[hptb!]
\centering
\includegraphics[width=6.5in]{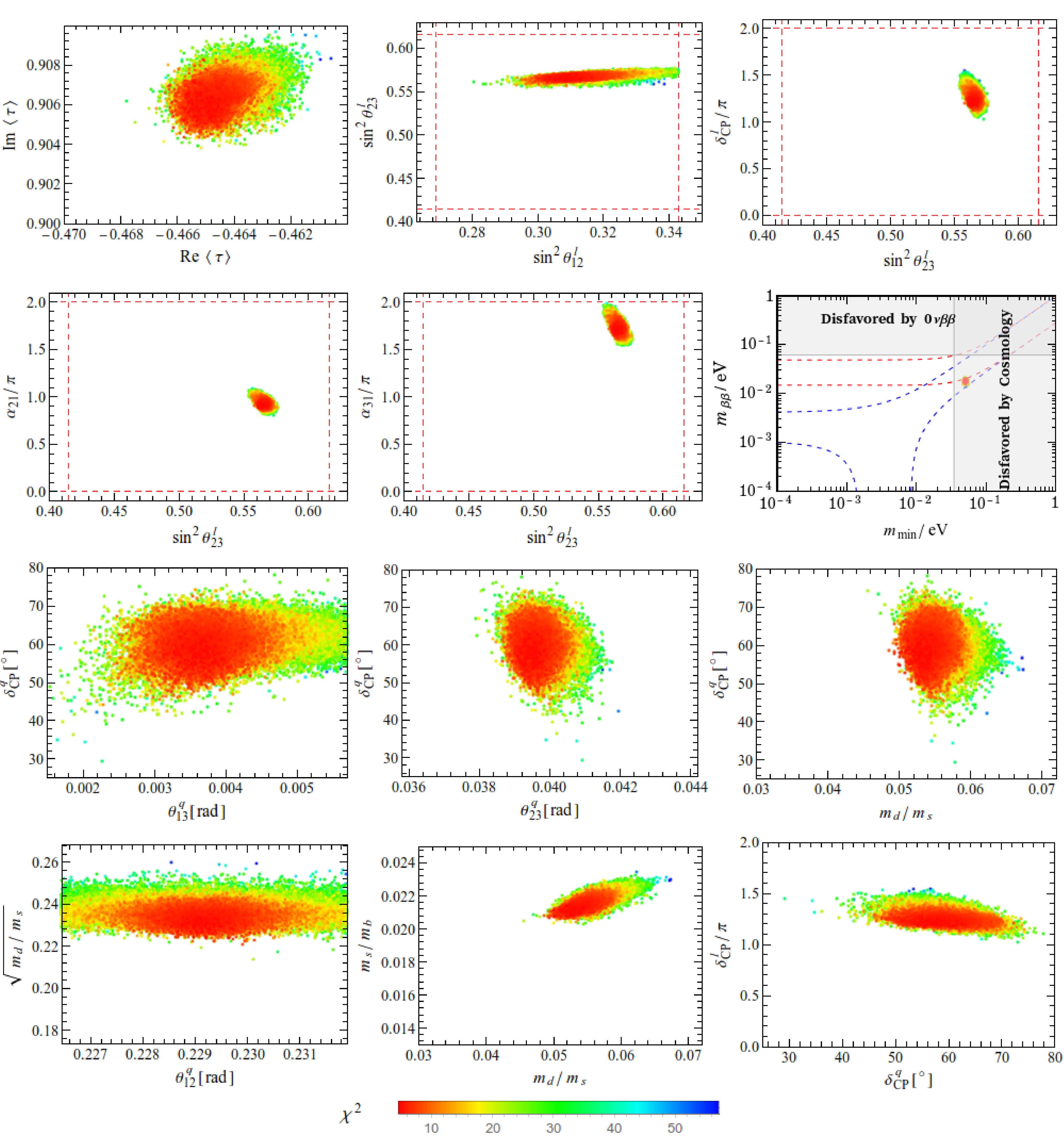}
\caption{The values of the complex modulus $\tau$ compatible with experimental data and the correlations between the neutrino mixing angles, CP violation phases, quark mass ratios and mixing parameters in the $\texttt{non-minimal model 3}$ with $\texttt{SS-II}$. The vertical and horizontal dashed lines are the $3\sigma$ bounds taken from~\cite{Esteban:2020cvm}. }
\label{fig:non_minimal_3_II}
\end{figure}

\begin{figure}[hptb!]
\centering
\includegraphics[width=6.5in]{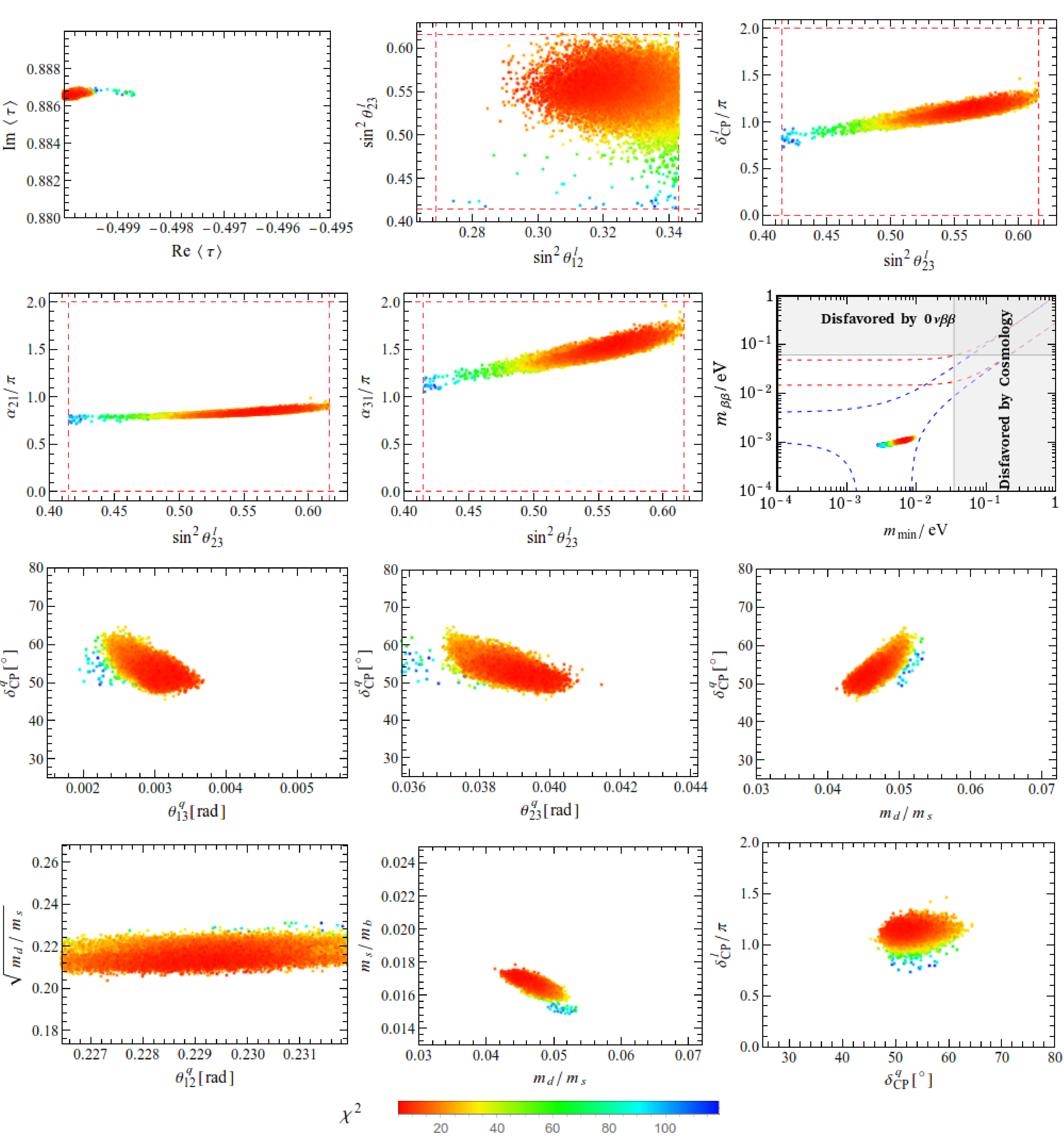}
\caption{The values of the complex modulus $\tau$ compatible with experimental data and the correlations between the neutrino mixing angles, CP violation phases, quark mass ratios and mixing parameters in the $\texttt{non-minimal model 3}$ with $\texttt{SS-I+II}$. The vertical and horizontal dashed lines are the $3\sigma$ bounds taken from~\cite{Esteban:2020cvm}. }
\label{fig:non_minimal_3_III}
\end{figure}

We display the allowed values of $\tau$ and the correlations among the fermion masses and mixing parameters for the $\texttt{non-minimal model 3}$ with type-I, type-II and type-I+II seesaw in figure~\ref{fig:non_minimal_3_I}, figure~\ref{fig:non_minimal_3_II} and figure~\ref{fig:non_minimal_3_III} respectively. From figure~\ref{fig:non_minimal_3_I} we see that the values of the complex modulus $\tau$ scatter in a small region close to the boundary of the fundamental region. The atmospheric mixing angle $\theta_{23}^{l}$ varies in the second octant and the allowed region of $\delta_{CP}^{l}$ is about $[ \pi,1.5\pi]$. The effective Majorana neutrino mass $m_{\beta\beta}$ is found to be about $1~\text{meV}$ with the lightest neutrino mass $m_1$ around $6~\text{meV}$. If the contribution of the type-II seesaw dominates over the other, $\theta_{23}^{l}$ is predicted to lie in the experimental favored $1\sigma$ region~\cite{Esteban:2020cvm}, as shown in figure~\ref{fig:non_minimal_3_II}. In the scenario of the mixed type-I and type-II seesaw mechanisms, from figure~\ref{fig:non_minimal_3_III} we can see that the allowed region of $\tau$ is close to the residual symmetry preserved point $-1/2+\sqrt{3}i/2$,
the atmospheric mixing angle $\theta_{23}^{l}$ is less constrained and its whole $3\sigma$ range can be covered, and the predicted value of $m_{\beta\beta}$ is around $1~\text{meV}$.

\section{Conclusion}
\label{section5}

In this paper we have combined $SO(10)$ Grand Unified Theories (GUTs) with $\Gamma_3\simeq A_4$ modular symmetry and presented a comprehensive analysis of the resulting quark and lepton mass matrices for all the simplest cases.
We have focussed on the case where the three fermion families in the 16 dimensional spinor representation form a triplet of
$\Gamma_3\simeq A_4$, with a Higgs sector comprising
a single Higgs multiplet $H$ in the ${\mathbf{10}}$ fundamental representation and one Higgs field $\overline{\Delta}$ in the ${\mathbf{\overline{126}}}$ for the minimal models, plus
an additional Higgs field $\Sigma$ in the ${\mathbf{120}}$ for the non-minimal models, all with specified modular weights.
The models are completely specified by the summation of the modular weights of matter fields and Higgs fields.
The neutrino masses are generated by the type-I and/or type II seesaw mechanisms and
results are presented for each model following an intensive numerical analysis where we have optimized the free parameters of the models in order to match the experimental data.
For the phenomenologically successful models, we present the best fit results in numerical tabular form as well as showing the most interesting graphical correlations between parameters, including leptonic CP phases and neutrinoless double beta decay, which have yet to be measured, leading to definite predictions for each of the models.

Once the modular weights are specified,
the Yukawa couplings are determined up to a number of overall dimensionless complex coefficients, and the value of the single complex modulus field $\tau$, which is the only flavon in the theory. All models with sums of modular weights up to 10 were considered,
and we presented results only for the simplest phenomenologically viable models.
We found that the minimal models containing only the Higgs fields in the ${\mathbf{10}}$ and the ${\mathbf{\overline{126}}}$ are the least viable, requiring both type I and type II seesaw mechanisms to be present simultaneously and also necessitating at least some sums of modular weights of 10 (if the sums of modular weights were restricted to 8 then no viable such models were found).
On the other hand, non-minimal models involving in addition to the Higgs fields in the ${\mathbf{10}}$ and the ${\mathbf{\overline{126}}}$, also a Higgs field in the ${\mathbf{120}}$, proved to be more successful, with many such models being found with sums of modular weights of up to 8 or less, and with the type I seesaw as well as the combined type I+II seesaw also being viable.
In the minimal models, more free parameters in the Yukawa couplings are necessary accommodate the experimental data of fermion masses and mixing, while fewer parameters are required in the Yukawa superpotential of non-minimal models, but the Higgs sector is more complicated due to the presence of $\Sigma$ in the representation $\mathbf{120}$. The Higgs potential required for the two light MSSM Higgs doublets to emerge from components of the ${\mathbf{10}}$, ${\mathbf{\overline{126}}}$ and $\mathbf{120}$ $SO(10)$ multiplets, leaving the other components heavy (the doublet-triplet splitting problem) was not considered.

In conclusion, we have successfully combined $SO(10)$ GUTs with $\Gamma_3\simeq A_4$ modular symmetry and analysed the simplest models of this kind, presenting the best fit results in numerical tabular form as well as showing the most interesting graphical correlations between parameters, leading to definite predictions for each of the models. The right-handed neutrino masses are predicted to be in the typical range for leptogenesis, and it would be interesting to study this in a future publication.

\section*{Acknowledgements}

We acknowledge Peng Chen for early participation in this project. JNL and GJD is supported by the National Natural Science Foundation of China under Grant Nos. 11975224, 11835013, 11947301 and the Key Research Program of the Chinese Academy of Sciences under Grant NO. XDPB15. SFK acknowledges the STFC Consolidated Grant ST/L000296/1 and the European Union's Horizon 2020 Research and Innovation programme under Marie Sk\l{}odowska-Curie grant agreement HIDDeN European ITN project (H2020-MSCA-ITN-2019//860881-HIDDeN).

\begin{table}[]
 \centering
 \begin{tabular}{|c|c|c|c|}
 \hline \hline
    \multirow{2}{*}{\texttt{Model}} & \texttt{Minimal Model 1} & \texttt{Minimal Model 2} & \texttt{Minimal Model 3}  \\
    \cline{2-4}
 &   \texttt{SS-I+II }&  \texttt{SS-I+II } &  \texttt{SS-I+II } \\   \hline$\tau$ & $0.463921+0.922876 i$ & $-0.332908+2.03892 i$ & $-0.135238+1.75306 i$\\
$r_2$ & $0.241839~e^{i 1.99966\pi}$ & $0.474468~e^{i 0.0214366\pi}$ & $0.481463~e^{i 0.00686157\pi}$\\
$\alpha_{1}v_L({\rm meV})$ & $373.984~e^{i 0.402224\pi}$ & $295.416~e^{i 0.0086878\pi}$ & $171.085~e^{i 0.00162031\pi}$\\
$\alpha_{2}/\alpha_{1}$ & $0.132025~e^{i 0.293718\pi}$ & $49.0324~e^{i 1.68401\pi}$ & $58.4926~e^{i 0.0188548\pi}$\\
$\alpha_{3}/\alpha_{1}$ & $0.386784~e^{i 0.0477546\pi}$ & $6.02243~e^{i 0.43811\pi}$ & $1.13999~e^{i 0.855299\pi}$\\
$\alpha_{4}/\alpha_{1}$ & $1.05696~e^{i 0.0249222\pi}$ & $14.1874~e^{i 0.69161\pi}$ & $59.3439~e^{i 0.0180549\pi}$\\
$\alpha_{5}/\alpha_{1}$ & $0.0656505~e^{i 1.29268\pi}$ & $191.441~e^{i 0.92892\pi}$ & $33.4546~e^{i 1.01527\pi}$\\
$\gamma_{1}/\alpha_{1}$ & $0.07221~e^{i 0.797625\pi}$ & $0.907953~e^{i 1.31986\pi}$ & $2.5552~e^{i 0.00204299\pi}$\\
$\gamma_{2}/\alpha_{1}$ & $0.00934649~e^{i 1.50211\pi}$ & $1.88799~e^{i 0.604461\pi}$ & $3.82826~e^{i 1.1549\pi}$\\
$\gamma_{3}/\alpha_{1}$ & $0.0673554~e^{i 0.806392\pi}$ & $22.1921~e^{i 0.792166\pi}$ & $1.08188~e^{i 1.04554\pi}$\\
$\gamma_{4}/\alpha_{1}$ & --- & $0.932598~e^{i 1.34128\pi}$ & $4.64699~e^{i 1.08188\pi}$\\
$\gamma_{5}/\alpha_{1}$ & --- & $0.131724~e^{i 0.216889\pi}$ & $8.35742~e^{i 0.00728868\pi}$\\
$\alpha_1 v_u^2/v_R({\rm meV})$ & $419.059$ & $92.8434$ & $51.4434$\\
$\alpha_1 v_u/{\rm GeV}$ & $30.1845$ & $3.4219$ & $1.6007$\\
$\alpha_1 r_1 v_d$/{\rm GeV} & $0.334326$ & $0.0374512$ & $0.0176275$\\
\hline $\sin^2\theta_{12}^l$ & $0.307937$ & $0.298492$ & $0.29805$\\
$\sin^2\theta_{13}^l$ & $0.0221603$ & $0.0221584$ & $0.0222692$\\
$\sin^2\theta_{23}^l$ & $0.569722$ & $0.576064$ & $0.578457$\\
$\delta_{CP}^l/^\circ$ & $190.306$ & $216.301$ & $233.631$\\
$\alpha_{21}/^\circ$ & $182.719$ & $197.457$ & $202.422$\\
$\alpha_{31}/^\circ$ & $168.88$ & $237.205$ & $262.258$\\
\hline $m_e/m_\mu$ & $0.00482513$ & $0.0046084$ & $0.00479302$\\
$m_\mu/m_\tau$ & $0.0587802$ & $0.0590124$ & $0.0593749$\\
$m_1/{\rm meV}$ & $6.04523$ & $4.91455$ & $5.14187$\\
$m_2/{\rm meV}$ & $10.5235$ & $9.9173$ & $10.0319$\\
$m_3/{\rm meV}$ & $50.4589$ & $50.4367$ & $50.5439$\\
$m_{\beta\beta}/{\rm meV}$ & $0.438542$ & $0.738522$ & $0.667982$\\
$M_1/{\rm GeV}$ & $7.49991\times 10^9$ & $1.3177\times 10^{10}$ & $8.90783\times 10^9$\\
$M_2/{\rm GeV}$ & $6.93057\times 10^{10}$ & $8.48068\times 10^{10}$ & $7.66469\times 10^{10}$\\
$M_3/{\rm GeV}$ & $1.06742\times 10^{12}$ & $3.72445\times 10^{11}$ & $3.39429\times 10^{11}$\\
$\alpha_{1}v_R/{\rm GeV}$ & $2.19591\times 10^{12}$ & $1.27381\times 10^{11}$ & $5.03052\times 10^{10}$\\
\hline $\theta_{12}^q$ & $0.229114$ & $0.229134$ & $0.229054$\\
$\theta_{13}^q$ & $0.00336806$ & $0.00278362$ & $0.0032868$\\
$\theta_{23}^q$ & $0.0401785$ & $0.0385602$ & $0.0400952$\\
$\delta_{CP}^q/^\circ$ & $65.6179$ & $61.712$ & $63.2177$\\
\hline $m_u/m_c$ & $0.00273513$ & $0.00265248$ & $0.00268673$\\
$m_c/m_t$ & $0.00275535$ & $0.00256208$ & $0.00252047$\\
$m_d/m_s$ & $0.0147522$ & $0.0479366$ & $0.0496554$\\
$m_s/m_b$ & $0.0216195$ & $0.0220083$ & $0.0204489$\\
\hline $m_b/m_\tau$ & $0.725858$ & $0.768468$ & $0.739121$\\
\hline $\chi^2_l$ & $0.252385$ & $1.68008$ & $2.27794$\\
$\chi^2_q$ & $31.1285$ & $5.27049$ & $1.23551$\\
$\chi^2_{b\tau}$ & $0.0190592$ & $1.6442$ & $0.0924274$\\
$\chi^2$ & $31.3999$ & $8.59476$ & $3.60588$\\
\hline $d_{\text{FT}}$ & $9.29068\times 10^6$ & $9.33133\times 10^6$ & $7.72748\times 10^6$\\
\hline \hline
\end{tabular}
\caption{\label{tab:fit-minimal_model}The best fit values of the free parameters and the corresponding predictions for the masses and mixing parameters of lepton and quark mixing at the best fit point in the $SO(10)$ minimal models characterized by the modular weights $(2k_{F}+k_{10},2k_F+k_{\overline{126}})=(10,6), (10,8), (10,10)$.
}
\end{table}

\begin{table}[]
 \centering
  \begin{tabular}{|c|c|c|}
\hline
\texttt{Non-minimal model} 1&   \texttt{SS-I } &  \texttt{SS-I+II }\\
\hline \hline$\tau$ & $-0.47503+0.897003 i$ & $0.492883+0.933623 i$\\
$r_2$ & $0.788723~e^{i 0.1\pi}$ & $0.79888~e^{i 0.0628905\pi}$\\
$r_3$ & $0.144759~e^{i 1.97672\pi}$ & $0.12476~e^{i 1.56739\pi}$\\
$c_e$ & $10.543~e^{i 0.380093\pi}$ & $13.1899~e^{i 0.304315\pi}$\\
$c_\nu$ & $8.91773~e^{i 0.456123\pi}$ & $11.0295~e^{i 0.457549\pi}$\\
$\alpha_{1}v_L({\rm meV})$ & --- & $39.1648~e^{i 0.000050371\pi}$\\
$\alpha_{2}/\alpha_{1}$ & $9.84853~e^{i 0.0687569\pi}$ & $1.93063~e^{i 0.245422\pi}$\\
$\alpha_{3}/\alpha_{1}$ & $9.84418~e^{i 0.0684131\pi}$ & $1.93307~e^{i 0.245233\pi}$\\
$\beta_{1}/\alpha_{1}$ & $0.717383~e^{i 0.480072\pi}$ & $0.11335~e^{i 1.05365\pi}$\\
$\beta_{2}/\alpha_{1}$ & $0.00999426~e^{i 1.26573\pi}$ & $0.0306156~e^{i 0.835143\pi}$\\
$\gamma_{1}/\alpha_{1}$ & $2.45837~e^{i 1.76308\pi}$ & $0.571433~e^{i 0.366949\pi}$\\
$\alpha_1 v_u^2/v_R({\rm meV})$ & $42.1872$ & $3.20875$\\
$\alpha_1 v_u/{\rm GeV}$ & $1.48805$ & $8.17528$\\
$\alpha_1 r_1 v_d$/{\rm GeV} & $0.0157628$ & $0.0864676$\\
\hline $\sin^2\theta_{12}^l$ & $0.30062$ & $0.294478$\\
$\sin^2\theta_{13}^l$ & $0.0223223$ & $0.0221286$\\
$\sin^2\theta_{23}^l$ & $0.574905$ & $0.569854$\\
$\delta_{CP}^l/^\circ$ & $221.268$ & $161.915$\\
$\alpha_{21}/^\circ$ & $193.648$ & $196.185$\\
$\alpha_{31}/^\circ$ & $223.403$ & $107.988$\\
\hline $m_e/m_\mu$ & $0.00480896$ & $0.00481101$\\
$m_\mu/m_\tau$ & $0.0591488$ & $0.0613664$\\
$m_1/{\rm meV}$ & $5.52462$ & $9.1019$\\
$m_2/{\rm meV}$ & $10.2333$ & $12.5317$\\
$m_3/{\rm meV}$ & $50.5706$ & $51.0373$\\
$m_{\beta\beta}/{\rm meV}$ & $0.020995$ & $1.92936$\\
$M_1/{\rm GeV}$ & $1.29033\times 10^{11}$ & $1.19024\times 10^{13}$\\
$M_2/{\rm GeV}$ & $1.29033\times 10^{11}$ & $1.19024\times 10^{13}$\\
$M_3/{\rm GeV}$ & $1.29033\times 10^{11}$ & $1.19024\times 10^{13}$\\
$\alpha_{1}v_R/{\rm GeV}$ & $5.30122\times 10^{10}$ & $2.10374\times 10^{13}$\\
\hline $\theta_{12}^q$ & $0.228946$ & $0.229003$\\
$\theta_{13}^q$ & $0.00417186$ & $0.00407841$\\
$\theta_{23}^q$ & $0.0409926$ & $0.041808$\\
$\delta_{CP}^q/^\circ$ & $43.6176$ & $49.7097$\\
\hline $m_u/m_c$ & $0.00278043$ & $0.00261272$\\
$m_c/m_t$ & $0.00248112$ & $0.00253582$\\
$m_d/m_s$ & $0.0537242$ & $0.0496242$\\
$m_s/m_b$ & $0.0188432$ & $0.0186282$\\
\hline $m_b/m_\tau$ & $0.820639$ & $0.782847$\\
\hline $\chi^2_l$ & $0.971826$ & $4.20832$\\
$\chi^2_q$ & $2.89378$ & $2.67676$\\
$\chi^2_{b\tau}$ & $9.12829$ & $3.10314$\\
$\chi^2$ & $12.9939$ & $9.98822$\\
\hline $d_{\text{FT}}$ & $1.14876\times 10^6$ & $1.38334\times 10^6$\\
\hline \hline
\end{tabular}
\caption{\label{tab:fit-non_minimal_model_1}The best fit values of the free parameters and the corresponding predictions for the masses and mixing parameters of lepton and quark mixing at the best fit point in the $SO(10)$ \texttt{non-minimal model 1} specified by the modular weights $(2k_F+k_{10}, 2k_F+k_{120}, 2k_F+k_{126})=(4, 8, 0)$.
}
\end{table}

\begin{table}[]
 \centering
  \begin{tabular}{|c|c|c|c|}
\hline
\texttt{Non-minimal model} 2&   \texttt{SS-I } & \texttt{SS-II }& \texttt{SS-I+II }\\
\hline \hline$\tau$ & $-0.499+0.890765 i$ & $-0.496235+0.935388 i$ & $0.498991+0.899733 i$\\
$r_2$ & $6.01436~e^{i 0.79951\pi}$ & $1.2475~e^{i 1.96305\pi}$ & $0.629183~e^{i 0.181499\pi}$\\
$r_3$ & $1.38203~e^{i 1.93548\pi}$ & $0.105474~e^{i 1.99953\pi}$ & $0.161103~e^{i 1.91134\pi}$\\
$c_e$ & $1.45966~e^{i 0.0999944\pi}$ & $10.0014~e^{i 1.08654\pi}$ & $6.4251~e^{i 1.37129\pi}$\\
$c_\nu$ & $0.888137~e^{i 0.597795\pi}$ & --- & $18.4103~e^{i 1.47163\pi}$\\
$\alpha_{1}v_L({\rm meV})$ & --- & $65.8371$ & $26.6754~e^{i 1.54135\pi}$\\
$\alpha_{2}/\alpha_{1}$ & $0.421745~e^{i 0.997015\pi}$ & $4.07592~e^{i 0.453719\pi}$ & $0.798097~e^{i 0.116969\pi}$\\
$\alpha_{3}/\alpha_{1}$ & $0.417481~e^{i 0.99881\pi}$ & $4.11381~e^{i 0.463642\pi}$ & $0.794603~e^{i 0.115025\pi}$\\
$\beta_{1}/\alpha_{1}$ & $0.517409~e^{i 0.000615941\pi}$ & $0.688854~e^{i 1.41383\pi}$ & $0.142793~e^{i 0.151608\pi}$\\
$\gamma_{1}/\alpha_{1}$ & $0.0601195~e^{i 1.04074\pi}$ & $3.17981~e^{i 0.575416\pi}$ & $0.488653~e^{i 0.537943\pi}$\\
$\gamma_{2}/\alpha_{1}$ & $0.0370351~e^{i 0.0111269\pi}$ & $0.0405593~e^{i 0.884191\pi}$ & $0.0841228~e^{i 1.36893\pi}$\\
$\gamma_{3}/\alpha_{1}$ & $0.0379525~e^{i 0.0150792\pi}$ & $0.0652665~e^{i 1.90002\pi}$ & $0.0762675~e^{i 1.38116\pi}$\\
$\alpha_1 v_u^2/v_R({\rm meV})$ & $6.71727$ & --- & $1.75721$\\
$\alpha_1 v_u/{\rm GeV}$ & $21.6382$ & $3.87595$ & $18.7023$\\
$\alpha_1 r_1 v_d$/{\rm GeV} & $0.364684$ & $0.0411592$ & $0.209795$\\
\hline $\sin^2\theta_{12}^l$ & $0.313701$ & $0.301997$ & $0.29938$\\
$\sin^2\theta_{13}^l$ & $0.0221311$ & $0.0220901$ & $0.0222548$\\
$\sin^2\theta_{23}^l$ & $0.561962$ & $0.540517$ & $0.577415$\\
$\delta_{CP}^l/^\circ$ & $203.802$ & $189.127$ & $195.014$\\
$\alpha_{21}/^\circ$ & $178.595$ & $301.399$ & $170.245$\\
$\alpha_{31}/^\circ$ & $245.394$ & $136.606$ & $236.075$\\
\hline $m_e/m_\mu$ & $0.0047742$ & $0.0047941$ & $0.00480169$\\
$m_\mu/m_\tau$ & $0.0586357$ & $0.0601214$ & $0.0605327$\\
$m_1/{\rm meV}$ & $8.39159$ & $66.9075$ & $5.74271$\\
$m_2/{\rm meV}$ & $12.0258$ & $67.4597$ & $10.3527$\\
$m_3/{\rm meV}$ & $50.7452$ & $83.6365$ & $50.6365$\\
$m_{\beta\beta}/{\rm meV}$ & $0.910213$ & $57.2755$ & $0.0677335$\\
$M_1/{\rm GeV}$ & $2.45575\times 10^{10}$ & --- & $3.28677\times 10^{12}$\\
$M_2/{\rm GeV}$ & $9.70305\times 10^{11}$ & --- & $2.09034\times 10^{13}$\\
$M_3/{\rm GeV}$ & $1.50727\times 10^{13}$ & --- & $9.13845\times 10^{13}$\\
$\alpha_{1}v_R/{\rm GeV}$ & $7.03997\times 10^{13}$ & --- & $2.01042\times 10^{14}$\\
\hline $\theta_{12}^q$ & $0.229066$ & $0.229088$ & $0.228868$\\
$\theta_{13}^q$ & $0.00330076$ & $0.00366392$ & $0.00378685$\\
$\theta_{23}^q$ & $0.0398076$ & $0.0405878$ & $0.0409927$\\
$\delta_{CP}^q/^\circ$ & $57.9865$ & $63.3747$ & $50.4869$\\
\hline $m_u/m_c$ & $0.00269002$ & $0.00270436$ & $0.0027786$\\
$m_c/m_t$ & $0.00261504$ & $0.00231404$ & $0.00247772$\\
$m_d/m_s$ & $0.0493918$ & $0.0576491$ & $0.0467914$\\
$m_s/m_b$ & $0.0206284$ & $0.0186664$ & $0.0193942$\\
\hline $m_b/m_\tau$ & $0.712848$ & $0.835678$ & $0.728889$\\
\hline $\chi^2_l$ & $1.10784$ & $3.11412$ & $0.864699$\\
$\chi^2_q$ & $1.27966$ & $2.46306$ & $1.34486$\\
$\chi^2_{b\tau}$ & $0.326885$ & $12.4086$ & $0.0013706$\\
$\chi^2$ & $2.71438$ & $17.9858$ & $2.21093$\\
\hline $d_{\text{FT}}$ & $1.28427\times 10^6$ & $1.61087\times 10^6$ & $1.47183\times 10^6$\\
\hline \hline
\end{tabular}
\caption{\label{tab:fit-non_minimal_model_2}
The best fit values of the free parameters and the corresponding predictions for the masses and mixing parameters of lepton and quark mixing at the best fit point in the $SO(10)$ \texttt{non-minimal model 2} specified by the modular weights $(2k_F+k_{10}, 2k_F+k_{120}, 2k_F+k_{\overline{126}})=(4, 2, 4)$. }
\end{table}

\begin{table}[]
 \centering
  \begin{tabular}{|c|c|c|c|}
\hline
\texttt{Non-minimal model} 3 &   \texttt{SS-I } & \texttt{SS-II }& \texttt{SS-I+II }\\
\hline \hline$\tau$ & $0.499999+0.897942 i$ & $-0.464246+0.906488 i$ & $-0.499998+0.886572 i$\\
$r_2$ & $1.40075~e^{i 0.17778\pi}$ & $1.35086~e^{i 0.041782\pi}$ & $7.50367~e^{i 0.845425\pi}$\\
$r_3$ & $0.211536~e^{i 1.97342\pi}$ & $0.126728~e^{i 1.90062\pi}$ & $1.97143~e^{i 1.88632\pi}$\\
$c_e$ & $2.99988~e^{i 0.590685\pi}$ & $12.2783~e^{i 1.70841\pi}$ & $1.35629~e^{i 1.75013\pi}$\\
$c_\nu$ & $6.63538~e^{i 1.97578\pi}$ & --- & $1.49696~e^{i 1.69818\pi}$\\
$\alpha_{1}v_L({\rm meV})$ & --- & $51.5461$ & $203.188~e^{i 1.61002\pi}$\\
$\alpha_{2}/\alpha_{1}$ & $3.38625~e^{i 1.27138\pi}$ & $4.39995~e^{i 0.789804\pi}$ & $0.537042~e^{i 0.983318\pi}$\\
$\alpha_{3}/\alpha_{1}$ & $3.35769~e^{i 1.2618\pi}$ & $4.57558~e^{i 0.799056\pi}$ & $0.543141~e^{i 0.981625\pi}$\\
$\beta_{1}/\alpha_{1}$ & $5.99313~e^{i 1.54363\pi}$ & $2.60037~e^{i 0.165305\pi}$ & $4.08986~e^{i 1.03948\pi}$\\
$\beta_{2}/\alpha_{1}$ & $0.143788~e^{i 0.1\pi}$ & $0.154082~e^{i 1.28314\pi}$ & $0.0076356~e^{i 1.94768\pi}$\\
$\gamma_{1}/\alpha_{1}$ & $5.64921~e^{i 0.969328\pi}$ & $3.79291~e^{i 0.785472\pi}$ & $0.109337~e^{i 0.976111\pi}$\\
$\gamma_{2}/\alpha_{1}$ & $2.83843~e^{i 0.943111\pi}$ & $0.0710498~e^{i 1.05365\pi}$ & $0.0522977~e^{i 1.99928\pi}$\\
$\gamma_{3}/\alpha_{1}$ & $2.83234~e^{i 0.951897\pi}$ & $0.094491~e^{i 1.88347\pi}$ & $0.0523395~e^{i 6.33014 \times 10^{-7}\pi}$\\
$\alpha_1 v_u^2/v_R({\rm meV})$ & $0.37705$ & --- & $9.85871$\\
$\alpha_1 v_u/{\rm GeV}$ & $2.05041$ & $3.3516$ & $14.8033$\\
$\alpha_1 r_1 v_d$/{\rm GeV} & $0.0285484$ & $0.0354997$ & $0.295418$\\
\hline $\sin^2\theta_{12}^l$ & $0.306759$ & $0.310936$ & $0.304737$\\
$\sin^2\theta_{13}^l$ & $0.0222446$ & $0.0222612$ & $0.0221234$\\
$\sin^2\theta_{23}^l$ & $0.547377$ & $0.566904$ & $0.567449$\\
$\delta_{CP}^l/^\circ$ & $233.574$ & $225.145$ & $219.734$\\
$\alpha_{21}/^\circ$ & $183.498$ & $166.982$ & $156.466$\\
$\alpha_{31}/^\circ$ & $325.281$ & $309.122$ & $295.647$\\
\hline $m_e/m_\mu$ & $0.00481232$ & $0.0048035$ & $0.0046923$\\
$m_\mu/m_\tau$ & $0.0617132$ & $0.0585641$ & $0.0580718$\\
$m_1/{\rm meV}$ & $6.3362$ & $50.5718$ & $7.104$\\
$m_2/{\rm meV}$ & $10.6933$ & $51.3002$ & $11.1654$\\
$m_3/{\rm meV}$ & $50.4152$ & $71.2635$ & $50.6344$\\
$m_{\beta\beta}/{\rm meV}$ & $0.912963$ & $17.8194$ & $1.10054$\\
$M_1/{\rm GeV}$ & $1.92962\times 10^{11}$ & --- & $1.56162\times 10^9$\\
$M_2/{\rm GeV}$ & $6.04195\times 10^{12}$ & --- & $4.19935\times 10^{11}$\\
$M_3/{\rm GeV}$ & $1.79534\times 10^{14}$ & --- & $6.74714\times 10^{12}$\\
$\alpha_{1}v_R/{\rm GeV}$ & $1.12617\times 10^{13}$ & --- & $2.24502\times 10^{13}$\\
\hline $\theta_{12}^q$ & $0.229089$ & $0.228991$ & $0.229058$\\
$\theta_{13}^q$ & $0.00360195$ & $0.00342953$ & $0.00327695$\\
$\theta_{23}^q$ & $0.0407079$ & $0.0397559$ & $0.0394391$\\
$\delta_{CP}^q/^\circ$ & $50.8017$ & $59.51$ & $53.0105$\\
\hline $m_u/m_c$ & $0.00269989$ & $0.0027034$ & $0.0027315$\\
$m_c/m_t$ & $0.00256261$ & $0.00245458$ & $0.00255892$\\
$m_d/m_s$ & $0.0503616$ & $0.0551891$ & $0.0451538$\\
$m_s/m_b$ & $0.0172515$ & $0.0216467$ & $0.0170418$\\
\hline $m_b/m_\tau$ & $0.74277$ & $0.769003$ & $0.704855$\\
\hline $\chi^2_l$ & $5.42448$ & $1.57742$ & $1.30864$\\
$\chi^2_q$ & $1.42362$ & $2.34845$ & $2.22642$\\
$\chi^2_{b\tau}$ & $0.1812$ & $1.69023$ & $0.702501$\\
$\chi^2$ & $7.0293$ & $5.6161$ & $4.23757$\\
\hline $d_{\text{FT}}$ & $2.72264\times 10^6$ & $2.99038\times 10^6$ & $3.03209\times 10^6$\\
\hline \hline
\end{tabular}
\caption{\label{tab:fit-non_minimal_model_3}The best fit values of the free parameters and the corresponding predictions for the masses and mixing parameters of lepton and quark mixing at the best fit point in the $SO(10)$ \texttt{non-minimal model 3} specified by the modular weights $(2k_F+k_{10}, 2k_F+k_{120}, 2k_F+k_{\overline{126}})=(4, 6, 4)$.}
\end{table}

\newpage


\providecommand{\href}[2]{#2}\begingroup\raggedright\endgroup

\end{document}